\documentclass[showpacs,floatfix,twocolumn,nofootinbib]{revtex4-1}
\usepackage{graphicx,psfrag,amsmath,amssymb,amsfonts,latexsym,color,epsf,dcolumn,graphpap}
\usepackage{subfig}
\usepackage{float}
\usepackage{lipsum}

\usepackage{tikz-cd} 
\usepackage{tikz}
\usetikzlibrary{shapes.geometric, arrows}

\definecolor{red}{rgb}{1,0,0}
\definecolor{blue}{rgb}{0,0,1}
\definecolor{skyblue}{rgb}{0,0,.5}
\definecolor{green}{rgb}{0,1,0}
\definecolor{orange}{cmyk}{0,.4,1,0}

\begin{document}
\title{Entanglement degradation of cavity modes due to the dynamical Casimir effect}
\author{Nicol\'as F.~Del Grosso, Fernando C.~Lombardo, Paula I.~Villar}
\affiliation{ Departamento de F\'\i sica {\it Juan Jos\'e Giambiagi}, FCEyN UBA and IFIBA CONICET-UBA, Facultad de Ciencias Exactas y Naturales, Ciudad Universitaria, Pabell\' on I, 1428 Buenos Aires, Argentina.\\}
\date{today}

\begin{abstract}
\noindent 
We study the entanglement dynamics between two cavities when one of them is harmonically shaken in the context of quantum information theory. We find four different regimes depending on the frequency of the motion and the spectrum of the moving cavity. If the moving cavity is three dimensional only two modes inside get coupled and the entanglement can either degrade asymptotically with time or oscillate depending on the driving. On the other hand, if the cavity has an equidistant spectrum the entanglement can either vanish asymptotically if it is driven with its fundamental frequency or have a sudden death if it is driven with an uneven harmonic frequency.

\end{abstract}
\maketitle
\section{Introduction}\label{sec:intro}

In the last decade a growing interest has arose in the field of relativistic quantum information (RQI) since  a number 
of interesting novel effects on entanglement between moving observers have been reported \cite{RQIresults}. Hence, this area of research has addressed questions related to the relativistic aspects of quantum physics, in particular submerging into the relation between the physics of quantum field theories and quantum information theory \cite{RQIreview}. As an example, black holes, once believed to be only mathematical artifacts, are today an established part of the universe and one the greatest mysteries of modern physics. They are very well defined in terms of general relativity but very little is known about their quantum nature. The discovery by Hawking \cite{Hawking_rad} that they emit thermal radiation and the information paradox that this fact had led to is one of the many efforts done in order to get a complete understanding of their relation with quantum mechanics. Particularly, RQI has a crucial role in  explaining quantum effects around black hole's physics. 

An interesting approach taken by authors in Ref.\cite{Fuentes2005} investigates the possibility of transmitting signals between inside and outside the event horizon of a black hole.  Therein, a pair of entangled particles is considered, each one hold by two different observers: one observer is inertial going into free-fall inside the black hole while the other  accelerates away and hovers outside the event horizon on a fixed distance. Authors noticed this study could be approximately described by  a scalar quantum field in Minkowski spacetime with two entangled modes. In this framework, they calculated the entanglement between two free modes of the scalar field as seen by an inertial observer detecting one of the modes and a uniformly accelerated observer detecting the second mode. They therefore stated  that as the acceleration of this observer increases, the perceived entanglement between the two modes  decreases asymptotically to zero, hindering the possibility of signal transmission \cite{Bouwmeester}. 
This work led to multiple ramifications and  numerous investigations on the effect of relativistic motion in quantum information \cite{Adesso2007,Friis2012R,Bruschi2012,Downes2011,Alsing2006,Lin2008}. For instance,  in Ref.\cite{Adesso2007}, the entanglement shared between observers in relative accelerated motion is analyzed finding that it vanishes between the lowest-frequency modes. In a different approach, authors in  Ref.\cite{Alsing2006}  showed that if the scalar field is replaced with a Dirac field the entanglement does not vanish in the infinite acceleration limit meaning that it can be used for quantum information tasks.  Sometimes,  results can be even more dramatic.
 In Ref.\cite{Lin2008} it has been shown  that the entanglement shared between two Unruh-DeWitt detectors 
(see Ref.\cite{Hu2012}) that are being accelerated can suddenly be lost in a finite amount of time. Similarly, in Refs.\cite{Bruschi2012, Friis2012} it has been stated that the entanglement shared by two entangled cavities in accelerated motion  oscillates as time elapses, during the time period the motion lasts. In this context, we can notice that  the original motivation has led to a more general question about how entanglement changes for observers in relative motion. In particular,  entanglement degradation caused by relativistic acceleration is a feature that has received great attention recently and we aim to  explore in this manuscript.

In this paper we will continue these investigations by studying how the entanglement between two cavities changes as one of them is rapidly shaken. In the previous cases, the accelerated motion of the observer changed the state of the field due to the Unruh effect \cite{Unruheffect}. Alternatively, in this manuscript,  the state of the field will be altered by the dynamical Casimir effect (DCE) \cite{DCE1}, known as being responsible for the creation of photons pairs from the electromagnetic vacuum when mirrors are subjected to ultra fast oscillations \cite{DCE2}. It is important to remark that, from an experimental point of view, making use of the Unruh effect requires either linearly unbounded motion which leads to short integration times or complex rotating setups \cite{Nature2017}. On the other hand for the DCE the motion is linear and bounded for arbitrarily long times of integration. Also, while creation of photon pairs out of quantum vacuum is close to being technologically accessible in optomechanical experiments \cite{Paraoanu, Ali, Nori, Wei, Nico} 
(practical optomechanical structures have been created in which the mirror can oscillate as fast as six billion times a second); it is already accessible in superconducting quantum circuits, where the DCE has been measured for the first time. Therefore practical implementations of the kind can be designed in order to get an insight into the consequences of relativistic motion on quantum entanglement.  This papers is organized as follows: in Section \ref{sec:DCE} we  review the physics of a cavity with two oscillating boundaries. We shall show that, in some cases, it can be reduced to the case of a cavity with a single moving mirror, allowing for an analytical solution  for small times. Then, in Sec. \ref{sec:EntanM} we  examine some properties of gaussian states in terms of the covariance matrix and see how we can measure the entanglement between two parts in that case. Sec. \ref{sec:Degrada} explores the  use of these tools to analytically study the entanglement degradation between two cavity modes due to the DCE. Finally, in Sec. \ref{sec:Conclu} we  present our conclusions.

\section{Dynamical Casimir effect}\label{sec:DCE}

We consider a rectangular cavity formed by  perfectly reflecting mirrors with dimensions $L_x$, $L_y$ and $L_z$ (we shall consider only $L_x$ in the case of a one-dimensional cavity). The mirrors oscillate together in the $x$-axis, maintaining the distance among them  fixed (we will refer to this cavity as a ``shaker"),  while the two other  pairs of mirrors in the $y$ and $z$-axis move so as to produce a rigid motion of the whole cavity \cite{DalvitMazzi2mirrors, pra95, 2squids}. We shall consider the light as a scalar field satisfying the wave equation
\begin{equation}
	\square \hat{\Phi}({\bf x}, t)=0
\end{equation}
subjected to time dependent boundary conditions on the mirrors with displacement $r(t)$, 
\begin{align}
\hat{\Phi}(x=0+r(t),y,z,t)&=\hat{\Phi}(x=L_x+r(t),y,z,t)=0\nonumber\\
\hat{\Phi}(x,y=0,z,t)&=\hat{\Phi}(x,y=L_y,z,t)=0\nonumber\\
\hat{\Phi}(x,y,z=0,t)&=\hat{\Phi}(x,y,z=L_z,t)=0.
\end{align}
This comes from the fact that TE and TM electromagnetic modes can be described by scalar fields \cite{Crocce2002}.

For $t<0$ all mirrors are at rest (static cavity) and the field can then be expanded as
\begin{equation}
\hat{\Phi}({\bf x},t)=\sum_{k}\left[\hat{a}_{k}^{\text{in}}u_{\bf k}({\bf x},t)+h.c.\right],
\label{eq:modos_in}
\end{equation}
where $\hat{a}_{n}^{\text{in}}$ are the bosonic operators corresponding to different photon modes and the functions $u_{\bf n}$ are the positive frequency solutions of the wave equation 
\begin{eqnarray}
u_{\bf k}({\bf x},t)&=&\sqrt{\frac{2}{L_x}}\sin(k_{x} x) \sqrt{\frac{2}{L_y}}\sin(k_{y}y)\sqrt{\frac{2}{L_z}}\sin(k_{z}z)\nonumber \\
&\times &\frac{e^{-i\omega_{\bf k} t}}{\sqrt{2\omega_{\bf k}}}
\label{eq:4}
\end{eqnarray}
with ${\bf k}= \left( \frac{n_x\pi}{L_x},\frac{n_y\pi}{L_y},\frac{n_z\pi}{L_z}\right)$ and $\omega_{\bf k}=|{\bf k}|$.
The mirrors in the $x$-axis start to move  at $t=0$ and the original basis gets continually deformed into a new one satisfying the boundary conditions $u_{\bf k}({\bf x},t)\to v_{\bf k}({\bf x},t)$. Expanding the field in this new basis as
\begin{equation}
\hat{\Phi}({\bf x},t)=\sum_{n}\left[\hat{a}_{n}^{\text{out}} v_{\bf n}({\bf x},t)+h.c.\right],
\label{eq:modos_out}
\end{equation}
we can define new bosonic operators $\hat{a}_{n}^{\text{out}}$ corresponding to a new notion of particles. As the mirrors go back to their original position and stop moving, the field can be expressed alternatively in any of the two basis, which means that there exist coefficients $\alpha_{\bf nk}$ and $\beta_{\bf nk}$ such that
\begin{equation}
	v_{\bf n}=\sum_{\bf k}  \left[\alpha_{\bf nk} u_{\bf k} +\beta_{\bf nk} u_{\bf k}^*  \right].
	\label{eq:seis}
\end{equation} 
Replacing this in Eq.(\ref{eq:modos_out}) and equating that to Eq.(\ref{eq:modos_in}) we can relate both sets of bosonic operators as 
\begin{equation}
\hat{a}_{\bf n}^{\text{out}}=\sum_{\bf k}  \left[\alpha_{\bf nk} \hat{a}_{\bf k}^{\text{in}} + \beta_{\bf nk}^* \hat{a}_{\bf k}^{\text{in}\dagger}  \right],
\label{eq:bogoliuvob}
\end{equation} 
which is known as a Bogoliubov transformation. In the following, we shall obtain explicit analytic solutions for the coefficients $\alpha_{\bf nk}$ and $\beta_{\bf nk}$ as a function of time for a harmonic oscillation of the cavity mirrors. This will give us the complete time evolution of the field in the Heisenberg picture.

It is important to remark that, for an initially vacuum field's  state,  the number of photons in mode ${\bf k}$ at a time $t>0$ can be computed as 
\begin{equation}
	\langle \hat{N}_{\bf n}\rangle=\langle \hat{a}_{\bf n}^{\text{out}\dagger}\hat{a}_{\bf n}^{\text{out}}\rangle=\sum_{\bf k} |\beta_{\bf nk}|^2.
	\label{eq:Nfot}
\end{equation}
This means that  for $|\beta_{\bf nk}|\neq 0$ photons  in mode ${\bf n}$ are created from vacuum. This is commonly known as  the Dynamical Casimir Effect (DCE).

\subsection*{Bogoliubov coefficients}
We shall focus on obtaining the  Bogoliubov coefficients. Hence, we begin by expanding solutions in the comoving basis
\begin{equation}
v_{\bf n}({\bf x},t)=\sum_{\bf k}Q_{\bf k}^{({\bf n})}(t)\varphi_{\bf k}(x,t)
\label{eq:Qk}
\end{equation}
where $Q_{\bf k}^{({\bf n})}(t)$ are time dependent coefficients and
\begin{eqnarray}
\varphi_{\bf k}({\bf x},t)&=&\sqrt{\frac{2}{L_x}}\sin(k_{x}(x-r(t))) \sqrt{\frac{2}{L_y}}\sin(k_{y}y) \nonumber \\
&\times & \sqrt{\frac{2}{L_z}}\sin(k_{z}z).
\label{eq:conmovil}
\end{eqnarray}

To find the time dependent coefficients $Q_{\bf k}^{({\bf n})}(t)$, we can replace Eq.(\ref{eq:Qk}) in Eq.(\ref{eq:modos_out}) and, then, this field decomposition into the wave equation. Finally, using the orthogonality of $\varphi_{\bf k}$

\begin{equation}
	\int_{r(t)}^{L+r(t)}\int_{0}^{L_y}\int_{0}^{L_z}\varphi_{\bf k}({\bf x})\varphi_{\bf j}({\bf x})dzdydx=\delta_{\bf kj}
\end{equation}
we find the following differential equation for the coefficients
\begin{align}
&\ddot{Q}_{\bf k}^{({\bf n})}(t)-2\dot{r}(t)\sum_{\bf j}\dot{Q}_{\bf j}^{({\bf n})}(t)g_{\bf kj}-\ddot{r}(t)\sum_{\bf j}Q_{\bf j}^{({\bf n})}(t)g_{\bf kj}\nonumber\\
&-\dot{r}^{2}(t)\sum_{\bf jl}Q_{\bf j}^{({\bf n})}(t)g_{\bf lj}g_{\bf lk}+\omega_{\bf k}^{2}Q_{\bf k}^{({\bf n})}(t)=0,
\label{eq:ddq}
\end{align}
where we have defined
\begin{align}
\label{eq:gkj}
g_{\bf kj}&=\int_{r(t)}^{L+r(t)}\int_{0}^{L_y}\int_{0}^{L_z}\partial\frac{\varphi_{\bf k}(x)}{\partial L_x}\varphi_{\bf j}(x)dzdydx\\
&=\begin{cases} \nonumber
\begin{array}{c}
\left((-1)^{j_{x}+k_{x}}-1\right)\frac{2k_{x}j_{x}}{k_{x}^{2}-j_{x}^{2}}\delta_{ k_yj_y}\delta_{k_zj_z}\\
0
\end{array} & \begin{array}{c}
k_{x}\neq j_{x}\\
k_{x}=j_{x}
\end{array}\end{cases}.
\end{align}

We will consider a harmonic oscillation of the mirrors displacement $r(t)=\epsilon\sin(\Omega t)$ and search for solutions of the form
\begin{equation}
Q_{\bf k}^{({\bf n})}=\alpha_{\bf n k}(\tau)\frac{e^{-i\omega_{\bf k}t}}{\sqrt{2\omega_{\bf k}}}+\beta_{\bf n k}(\tau)\frac{e^{i\omega_{\bf k}t}}{\sqrt{2\omega_{\bf k}}}
\label{eq:QAB}
\end{equation}
in order to get  analytical predictions about the particle creation process (this is a standard procedure known as {\it multiple scale analysis} \cite{3DDCE}). Hence, if we  introduce a slow time $\tau:=\frac{1}{2L_x}\epsilon \omega_1 t$ (with $\omega_1$ the fundamental frequency defined after Eq. \ref{eq:4}),  the functions $\alpha_{\bf n k}$ and $\beta_{\bf n k}$  are  slowly varying and contain the cumulative resonant effects. These functions will have initial conditions $\alpha_{\bf n k}(\tau=0)=\delta_{\bf n k}$ and $\beta_{\bf n k}(\tau=0)=0$ so that the initial field is given by Eq. (\ref{eq:modos_in}). After substituting Eq. (\ref{eq:QAB}) in eq. (\ref{eq:ddq}) and averaging  over fast oscillations we obtain the following differential equations 
\begin{widetext}
	\begin{align}
	\frac{d\beta_{\bf n k}}{d\tau}&=\sum_{\bf j}\frac{\Omega}{2\omega_{\bf k}}g_{\bf kj}\beta_{\bf n j}\left[\left(\omega_{\bf j}+\frac{\Omega}{2}\right)\delta(\Omega+\omega_{\bf j}-\omega_{\bf k})+\left(\omega_{\bf j}-\frac{\Omega}{2}\right)\delta(-\Omega+\omega_{\bf j}-\omega_{\bf k})\right]\nonumber\\
	&+\sum_{\bf j}\frac{\Omega}{2\omega_{\bf k}}g_{\bf kj}\left(-\omega_{\bf j}+\frac{\Omega}{2}\right)\alpha_{\bf n j}\delta(\Omega-\omega_{\bf j}-\omega_{\bf k})
	\label{eq:A}
	\end{align}
	\begin{align}
	\frac{d\alpha_{\bf n k}}{d\tau}&=\sum_{\bf j}\frac{\Omega}{2\omega_{\bf k}}g_{\bf kj}\alpha_{\bf n j}\left[\left(\omega_{\bf j}+\frac{\Omega}{2}\right)\delta(\Omega+\omega_{\bf j}-\omega_{\bf k})+\left(\omega_{\bf j}-\frac{\Omega}{2}\right)\delta(-\Omega+\omega_{\bf j}-\omega_{\bf k})\right]\nonumber\\
	&+\sum_{\bf j}\frac{\Omega}{2\omega_{\bf k}}g_{\bf kj}\left(-\omega_{\bf j}+\frac{\Omega}{2}\right)\beta_{\bf n j}\delta(\Omega-\omega_{\bf j}-\omega_{\bf k}),
	\label{eq:B}
	\end{align}
\end{widetext}
where $\delta(0)=1$ and $\delta(x)=0$ for $x\neq0$.

 We can mention that by replacing Eqs.(\ref{eq:QAB}) and (\ref{eq:conmovil}) in Eq.(\ref{eq:Qk}) and comparing the result with that of Eq.(\ref{eq:seis}), it is easy to note  that these slowly changing $\alpha_{\bf n k}$ and $\beta_{\bf n k}$ will be the Bogoliubov coefficients found in the previous section.

 There are essentially two distinct cases in which we can solve this set of coupled differential equations depending on whether the cavity's spectrum is equidistant or non-equidistant. As for a non-equidistant spectrum, it can be achieved for example by considering a three-dimensional cavity or, in another case by  considering a massive scalar field. In such  cases, only two modes will couple (except for some exceptional cases described in  \cite{Crocce2002}). Oppositely, if the cavity is one-dimensional; the spectrum is equidistant $\omega_n=n\pi/L_x$ and then infinitely many modes will couple.  In the following, we shall briefly describe both situations.

\subsubsection*{3D cavity and non-equidistant spectrum: Two coupled modes}

Let us consider the case in which  we have a three dimensional cavity. If we assume a driving (shaking) frequency $\Omega$ that can be formed by two modes $\omega_{\bf s}$ and $\omega_{\bf c}$, as $\Omega=\omega_{\bf s}+\omega_{\bf c}$; then, because the spectrum is non-equidistant, there is (almost) never a third mode $\omega_{\bf d}$ such that $\Omega=|\omega_{\bf s}\pm\omega_{\bf d}|$ or $\Omega=|\omega_{\bf c}\pm\omega_{\bf d}|$. In that case, Eqs.(\ref{eq:A}) and (\ref{eq:B}) reduce to
\begin{align}
\frac{d\beta_{\bf n s}}{d\tau}&=\frac{\Omega}{2\omega_{\bf s}}g_{\bf sc}\left(-\omega_{\bf c}+\frac{\Omega}{2}\right)\alpha_{\bf n c}\\
\frac{d\alpha_{\bf n c}}{d\tau}&=\frac{\Omega}{2\omega_{\bf c}}g_{\bf cs}\left(-\omega_{\bf s}+\frac{\Omega}{2}\right)\beta_{\bf n s}.
\end{align}
This is a linear system with four coupled differential equations that can be easily solved.  When combined with 
Eq.(\ref{eq:bogoliuvob}) we obtain
\begin{align}
a_{\bf s}^{\text{out}}&=\cosh(\gamma^-\tau)a_{\bf s}^{\text{in}}+\sinh(\gamma^-\tau)a_{\bf c}^{\text{in}\dagger} \label{eq:sumas}\\ 
a_{\bf c}^{\text{out}}&=\cosh(\gamma^-\tau)a_{\bf c}^{\text{in}}+\sinh(\gamma^-\tau)a_{\bf s}^{\text{in}\dagger}\label{eq:sumap}
\end{align}
where
\begin{equation}
\gamma^-=\frac{\Omega}{2}g_{\bf sc}\sqrt{\frac{\left(\omega_{\bf s}-\frac{\Omega}{2}\right)\left(\frac{\Omega}{2}-\omega_{\bf c}\right)}{\omega_{\bf s}\omega_{\bf c}}}.
\end{equation}
This transformation generates a two-mode squeeze state with squeezing parameter
$\gamma^-\tau$, occurring when a driving quanta is converted into a pair of photons in modes $\omega_{\bf s}$ and $\omega_{\bf c}$. It is important to remark that this fact can also be seen from our previous discussion. The number of photons created from the vacuum is given by Eq.(\ref{eq:Nfot}), and since, in this case it is $\beta_{\bf nk}\neq0$, there is creation of particles and it is exponential in time  \cite{3DDCE,Num1}.


The other possible coupling between two modes occurs if $\Omega=|\omega_{\bf s}-\omega_{\bf c}|$.  In that case, the system of Eq. (\ref{eq:B}) becomes 
\begin{align}\frac{d\alpha_{\bf n s}}{d\tau}&=\frac{\Omega}{2\omega_{\bf s}}g_{\bf sc}\left(\omega_{\bf c}-\frac{\Omega}{2}\right) \alpha_{\bf n c}\\
\frac{d\alpha_{\bf n c}}{d\tau}&=\frac{\Omega}{2\omega_{\bf c}}g_{\bf cs}\left(\omega_{\bf s}+\frac{\Omega}{2}\right) \alpha_{\bf n s}.
\end{align}
This set of equations can be readily solved and yields 
\begin{align}
a_{\bf s}^{\text{out}}&=\cos(\gamma^+\tau)a_{\bf s}^{\text{in}}+\sin(\gamma^+\tau)a_{\bf c}^{\text{in}} \label{eq:restas}\\ 
a_{\bf c}^{\text{out}}&=\cos(\gamma^+\tau)a_{\bf c}^{\text{in}}-\sin(\gamma^+\tau)a_{\bf s}^{\text{in}}.
\label{eq:restac}
\end{align}
where 
\begin{equation}
\gamma^+=\frac{\Omega}{2}g_{\bf sc}\sqrt{\frac{\left(\omega_{\bf s}+\frac{\Omega}{2}\right)\left(\omega_{\bf c}-\frac{\Omega}{2}\right)}{\omega_{\bf s}\omega_{\bf c}}}.
\end{equation}
It is worth mentioning that in this situation $\beta_{\bf nk}=0$ and no new photons are created. This means  that for an initially vacuum state the dynamics is trivial. However, given an initial state with some photons in one of the modes, the number of photons in each mode will oscillate while keeping the total number of photons constant.

\subsubsection*{1D cavity and equidistant spectrum: Infinite coupled modes}

We now consider the case where the cavity is one dimensional \cite{1DDCE} and the resulting spectrum is equidistant. Since the only relevant direction corresponds to the $x$-axis we discard the bold face indices to keep only this component. Then, using the coefficients described in Eq.(\ref{eq:gkj}), the set of Eqs. (\ref{eq:A}) and (\ref{eq:B}) reduces to 
\begin{widetext}
	\begin{align}
	\frac{d\beta_{ n k}}{d\tau}&=\sum_{ j+k\text{ odd}}\frac{\Omega}{2\omega_{ k}}g_{ kj}\beta_{ n j}\left[\left(\omega_{ j}+\frac{\Omega}{2}\right)\delta(q+ j-k)+\left(\omega_{ j}-\frac{\Omega}{2}\right)\delta(-q+j- k)\right]\nonumber\\
	&+\sum_{j+k\text{ odd}}\frac{\Omega}{2\omega_{ k}}g_{ kj}\left(-\omega_{ j}+\frac{\Omega}{2}\right)\alpha_{ n j}\delta(q- j- k)
	\label{eq:A1}
	\end{align}
	\begin{align}
	\frac{d\alpha_{ n k}}{d\tau}&=\sum_{ j+k\text{ odd}}\frac{\Omega}{2\omega_{ k}}g_{ kj}\alpha_{ n j}\left[\left(\omega_{ j}+\frac{\Omega}{2}\right)\delta(q+j-k)+\left(\omega_{ j}-\frac{\Omega}{2}\right)\delta(-q+ j- k)\right]\nonumber\\
	&+\sum_{ j+k\text{ odd}}\frac{\Omega}{2\omega_{ k}}g_{ kj}\left(-\omega_{ j}+\frac{\Omega}{2}\right)\beta_{ n j}\delta(q- j- k),
	\label{eq:B1}
	\end{align}
\end{widetext}
where we have defined $q=\Omega/\omega_1$ and used the fact  that the spectrum is given by $\omega_{ j}=j\omega_1$ and that $g_{kj}=0$ if $k+j$ is even. It is important to note that, since $k+j$ is odd, $k-j$ is also odd. This means that if $q$ is not an odd number then the RHS of Eqs.(\ref{eq:A1}) and (\ref{eq:B1}) is null and the evolution is trivial.

In order to obtain an  analytical solution for the system of Eqs. (\ref{eq:A1}) and (\ref{eq:B1}), we consider the situation in which only one wall (let say the right one) of the cavity oscillates,  in the one dimensional case. In such a case, a derivation similar  to the one performed in the previous section leading to Eqs. (\ref{eq:A}-\ref{eq:B}), leads to the following system of equations
\begin{widetext}
	\begin{align}
	\frac{d\tilde{\beta}_{ k}^{({ n})}}{d\tau}&=-\frac{\pi^2k^2}{2\omega_{k}L_x^2}\tilde{\alpha}^{({ n})}_{ k}\delta(2 k-q)+\sum_{ j}\frac{\Omega}{2\omega_{ k}}\tilde{g}_{ kj}\tilde{\beta}_{ j}^{({ n})}\left[\left(\omega_{ j}+\frac{\Omega}{2}\right)\delta(q+ j-k)+\left(\omega_{ j}-\frac{\Omega}{2}\right)\delta(-q+j- k)\right]\nonumber\\
	&+\sum_{j}\frac{\Omega}{2\omega_{ k}}\tilde{g}_{ kj}\left(-\omega_{ j}+\frac{\Omega}{2}\right)\tilde{\alpha}_{ j}^{({ n})}\delta(q- j- k)
	\label{eq:A2}
	\end{align}
	\begin{align}
	\frac{d\tilde{\alpha}_{ k}^{({ n})}}{d\tau}&=-\frac{\pi^2k^2}{2\omega_{ k}L_x^2}\tilde{\beta}^{({\bf n})}_{ k}\delta(2 k-q)+\sum_{ j}\frac{\Omega}{2\omega_{ k}}\tilde{g}_{\bf kj}\tilde{\alpha}_{ j}^{({ n})}\left[\left(\omega_{ j}+\frac{\Omega}{2}\right)\delta(q+j-k)+\left(\omega_{ j}-\frac{\Omega}{2}\right)\delta(-q+ j- k)\right]\nonumber\\
	&+\sum_{ j}\frac{\Omega}{2\omega_{ k}}\tilde{g}_{ kj}\left(-\omega_{ j}+\frac{\Omega}{2}\right)\tilde{\beta}_{ j}^{({ n})}\delta(q- j- k),
	\label{eq:B2}
	\end{align}
\end{widetext}
with 
\begin{align}
\tilde{g}_{ kj}=\begin{cases} 
\begin{array}{c}
\frac{2kj}{k^{2}-j^{2}}\\
0
\end{array} & \begin{array}{c}
k\neq j\\
k=j .
\end{array}\end{cases}
\end{align}
This set of equations is very similar to Eqs. (\ref{eq:A1}) and (\ref{eq:B1}), the differences being only two. The first difference among the systems of equations is that  Eqs. (\ref{eq:A2}) and (\ref{eq:B2}) have an extra term proportional to $\delta(2k-q)$ (which is not present in  Eqs. (\ref{eq:A2}) and (\ref{eq:B2})) \cite{Num2}. This term creates pair of photons in mode ${k}$ and is only relevant when $q=2 k$. The second difference is that $\tilde{g}_{ kj}\neq g_{ kj}$. However, it is easy to see that $2\tilde{g}_{ kj}=g_{ kj}$ if $k+j$ is odd, and this factor 2 can be absorbed in the oscillation amplitude. 
Therefore, if we excite a cavity with one moving wall at an odd frequency $q$, we get that the first term of RHS of 
Eqs.(\ref{eq:A2}) and (\ref{eq:B2}) vanishes. All others also disappear, except for the case of $k+j$ being odd (this is due to the $\delta$ factors).    
This proves that when  $q$ is odd Eqs. (\ref{eq:A1}) and (\ref{eq:B1}) are equivalent to Eqs. (\ref{eq:A2}) and (\ref{eq:B2}). Consequently,  this means that a cavity being shaken rigidly with an odd frequency behaves just as an identical cavity with only one wall oscillating with the same frequency and double amplitude. This result is very important and, although a relationship had been noticed \cite{mazzi2}, it had not been stated in this way in the existing literature. We shall take advantage of this equivalence to get analytical predictions on the particle creation process since analytical solutions for the system with one moving wall are well known \cite{Dodonov1998}. In such a case the out operators are given by 
\begin{equation}
a_{m}^{\text{out}}=\sum_{n=1}^{\infty}\sqrt{\frac{m}{n}}\left[\rho^{(n)}_m a_{m}^{\text{in}}-\rho^{(n)*}_{-m} a_{m}^{\text{in}\dagger}\right].\label{eq:1d}
\end{equation}
where 
\begin{align}
\rho^{(j+nq)}_{j+mq}&=\frac{\Gamma(1+n+j/q)(\sigma \kappa)^{n-m}}{\Gamma(1+m+j/q)\Gamma(1+n-m)}\nonumber\\
&\times F(n+j/q,-m-j/q;1+n-m;\kappa^2)
\label{eq:coef1}
\end{align}
for $j=0,1,...,q$, $n=1,2,3,...$ $m=\pm1,\pm2,\pm3,...$ and
\begin{align}
\rho^{(k+nq)}_{j+mq}&=0
\label{eq:coef2}
\end{align}
if $j\neq k$, where
\begin{align}
\sigma&=(-1)^q\\
\kappa&=\tanh(q \tilde{\tau}).
\end{align}
As it has been exposed above, these are also solutions of the rigidly shaken cavity taking $\tilde{\tau}=2\tau=\epsilon\omega_1 t$, that is doubling the amplitude $\epsilon$. 

Similarly to what we have noticed before in the three dimensional case, there are two qualitatively different regimes depending on the driving frequency. If $\Omega=\omega_1=\omega_{j+1}-\omega_j$ then no photons are created as a whole. However, given an initial number they can ``jump'' between adjacent modes. For long times, the number of photons on any mode goes to zero as they are lost to higher frequency modes. Alternatively, if $\Omega=q \omega_1$ with $q$ an odd number greater than 1, pairs of photons are created from the vacuum on any pair of modes $j$ and $k$ such that $\Omega=\omega_j+\omega_k$.

We have briefly analyzed  DCE in different configurations. In all cases, by shaking the cavity, we 
have altered the state of the scalar field.  In addition, we have shown that, in some particular cases,  we obtain creation of particles
as a further result. In the following we shall study the entanglement process between two of these cavities as one of them is rapidly shaken.

 \begin{figure*}
	\includegraphics[width=.9\linewidth]{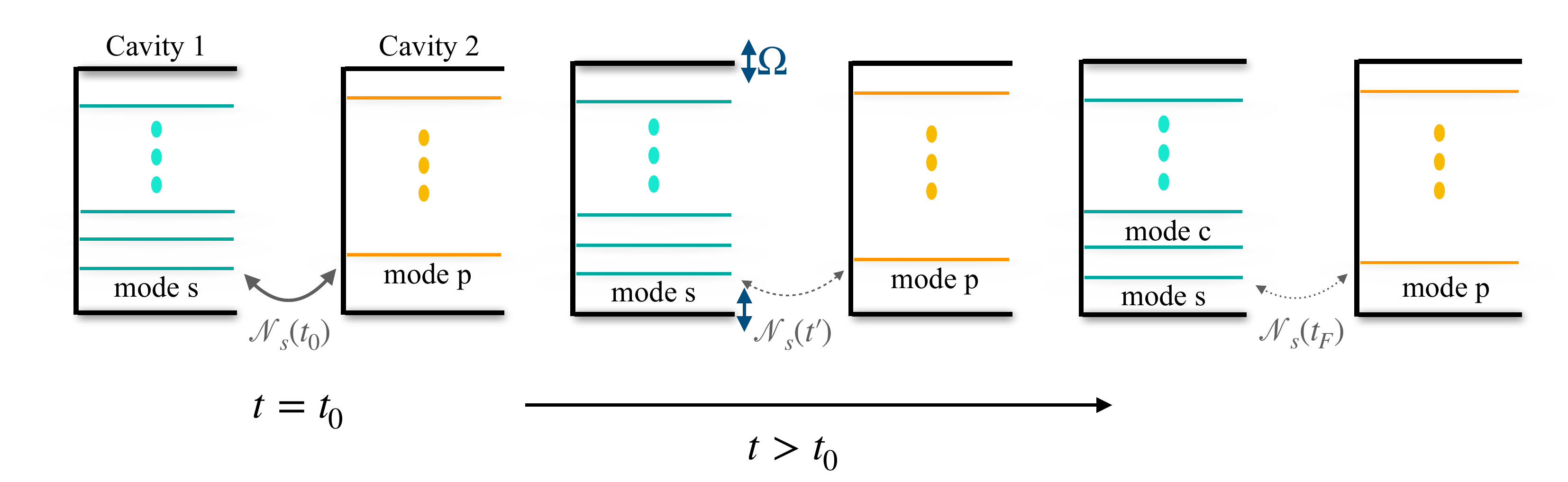}
	\caption{The two cavities are initially (at $t = t_0$) static and modes $s$ and $p$ are entangled (thick gray line). As we start shaking cavity 1 at time $t = t'$ with frequency $\Omega$,  the entanglement between these modes becomes weaker. (dashed gray line). Finally, at $t_F$ the shaking stops but the entanglement (measured by negativity ${\cal N}_s$), is much weaker than initially ${\cal N}_s(t_F) < {\cal N}_s(t_0)$ (dotted line).}
	\label{fig:esquema}
\end{figure*}
\section{Entanglement measurement}\label{sec:EntanM}

A particularly relevant set of states, from a theoretical viewpoint, are gaussian states; which include coherent, squeezed, and thermal states. Given a set of harmonic oscillators, we can take a basis of quadrature operators $R=(q_1,p_1,...,q_n, p_n)$, where 
\begin{align}
q_j&=\frac{1}{\sqrt{2}}(a_j+a^\dagger_j) \label{eq:quadrature1}\\
p_j&=\frac{-i}{\sqrt{2}}(a_j-a^\dagger_j)
\label{eq:quadrature2}
\end{align}
and completely characterize a gaussian state $\rho$ with its displacement vector 
\begin{equation}
	d_j=\langle R_j \rangle_{\rho}
\end{equation}
and its covariance matrix 
\begin{equation}
V_{ij}=\frac{1}{2}\langle R_i R_j+R_j R_i \rangle_{\rho}- \langle R_i \rangle_{\rho} \langle R_j \rangle_{\rho}.
\label{eq:covariance}
\end{equation}
Some important properties of these states are that the evolution of a gaussian state under a quadratic hamiltonian is also a gaussian state. Further, given a system in a globally gaussian state, any subsystem has a state that is also gaussian and its covariance matrix is given by the restriction of the covariance matrix of the whole to that subsystem.  In addition, it is possible to
quantify the entanglement in a mixed state \cite{entaglemix} by using the logarithmic negativity \cite{lognegat}, which is an entanglement monotone given by 
\begin{equation}
	\mathcal{N}(\rho)=\log_2||\rho^{\Gamma_A}||_1,
\end{equation}
where $||\rho^{\Gamma_A}||_1$ is the trace norm of the partial transpose of $\rho$ with respect to subsystem $A$. 

Herein, we want to calculate the entanglement between two modes in gaussian states. In such case, the covariance matrix is of the form
\begin{equation}
	V_{A|B}=\left|\begin{array}{cc}
	V_A & V_C\\
	V_C^{T} & V_B
	\end{array}\right|
\end{equation} 
and the logarithmic negativity can be calculated as \cite{lognegat2}
\begin{equation}
\mathcal{N}=\max\{0, -\log 2\nu_-\}
\label{eq:logneg}
\end{equation} 
where 
\begin{equation}
\nu_{-}=\sqrt{\frac{\Sigma}{2}-\frac{1}{2}\sqrt{\Sigma^{2}-4\det V_{A|B}}}
\end{equation} 
and 
\begin{equation}
	\Sigma=\det V_A+\det V_B-2\det V_C.
\end{equation}

In order to determine if there is still information shared between subsystems when the entanglement vanishes, we will 
 use the mutual information \cite{mutualInf1,mutualInf2}
\begin{equation}
I(\rho_{A|B})=S_V(\rho_A)+S_V(\rho_B)-S_V(\rho_{A|B}),
\end{equation}
which measures the total correlations (quantum and classical), with $S_V$ being the Von Neumann entropy.
In the case of a gaussian state, it can be easily computed as 
\begin{eqnarray}
I(V_{A|B})&=&f(\sqrt{\det 2 V_A})+f(\sqrt{\det 2 V_B}) \nonumber \\
&-&f(\eta^-_{A|B})-f(\eta^+_{A|B})
\end{eqnarray}
with
\begin{equation}
f(x)=\frac{x+1}{2}\log\left(\frac{x+1}{2}\right)-\frac{x-1}{2}\log\left(\frac{x-1}{2}\right)
\end{equation}
where ${\eta^-_{A|B},\eta^+_{A|B}}$ are the symplectic eigenvalues of $2V_{A|B}$.\\

 \begin{figure*}
	\subfloat[\label{sfig:prod_vm}]{%
		\includegraphics[width=.33\linewidth]{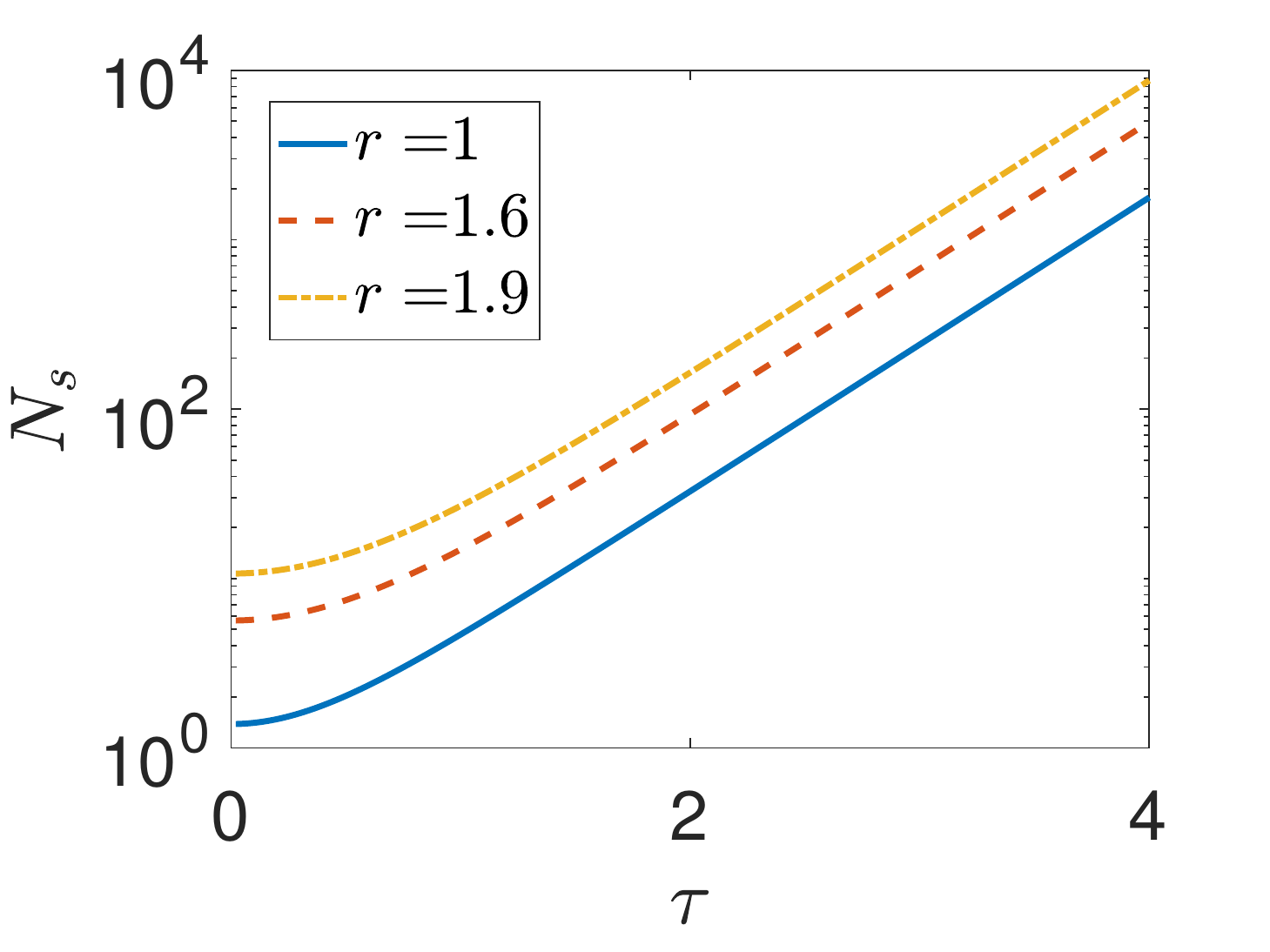}%
	}\hfill
	\subfloat[\label{sfig:prod_eta}]{%
		\includegraphics[width=.33\linewidth]{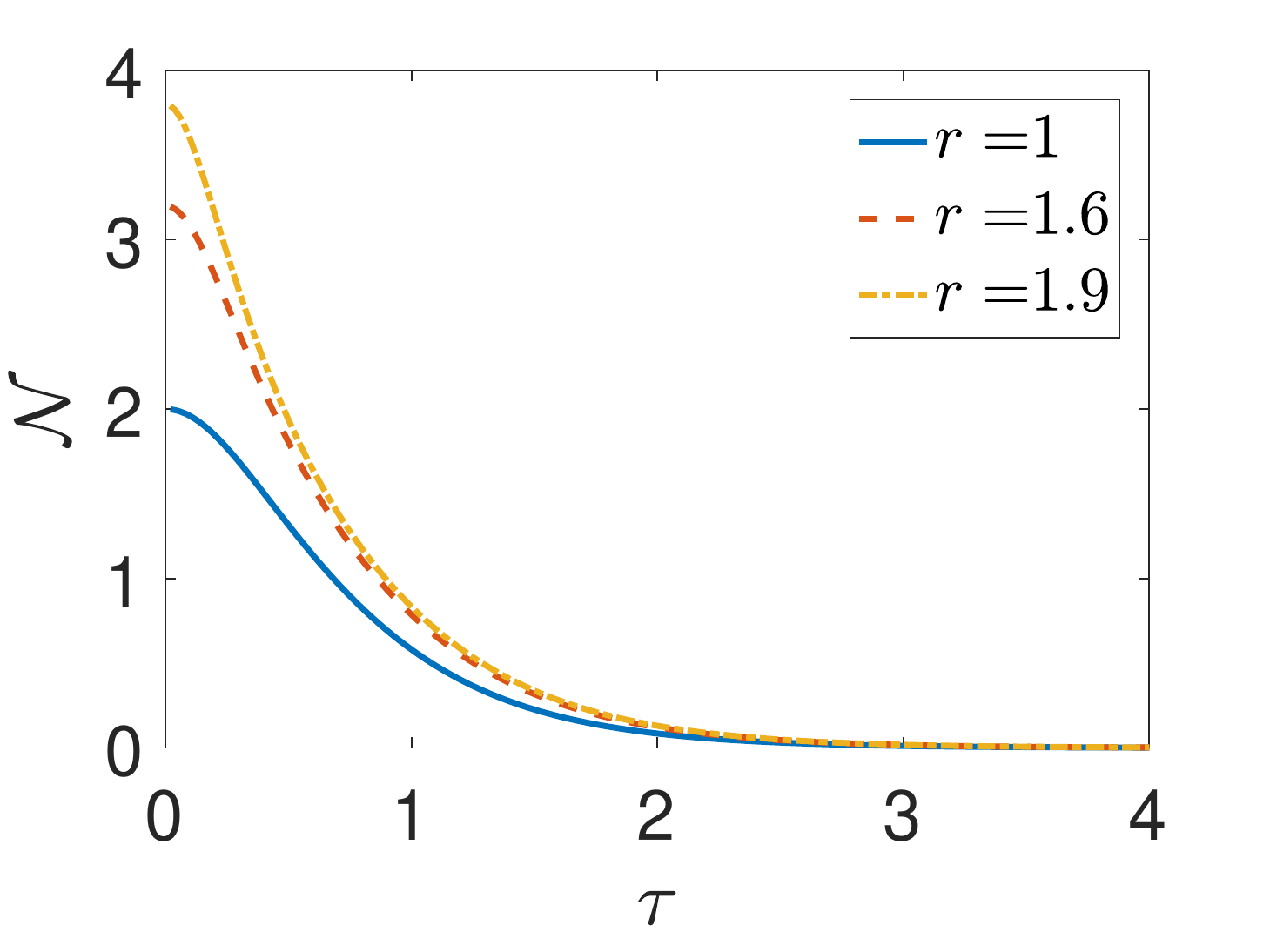}%
	}\hfill
	\subfloat[\label{sfig:prod_s_t}]{%
		\includegraphics[width=.33\linewidth]{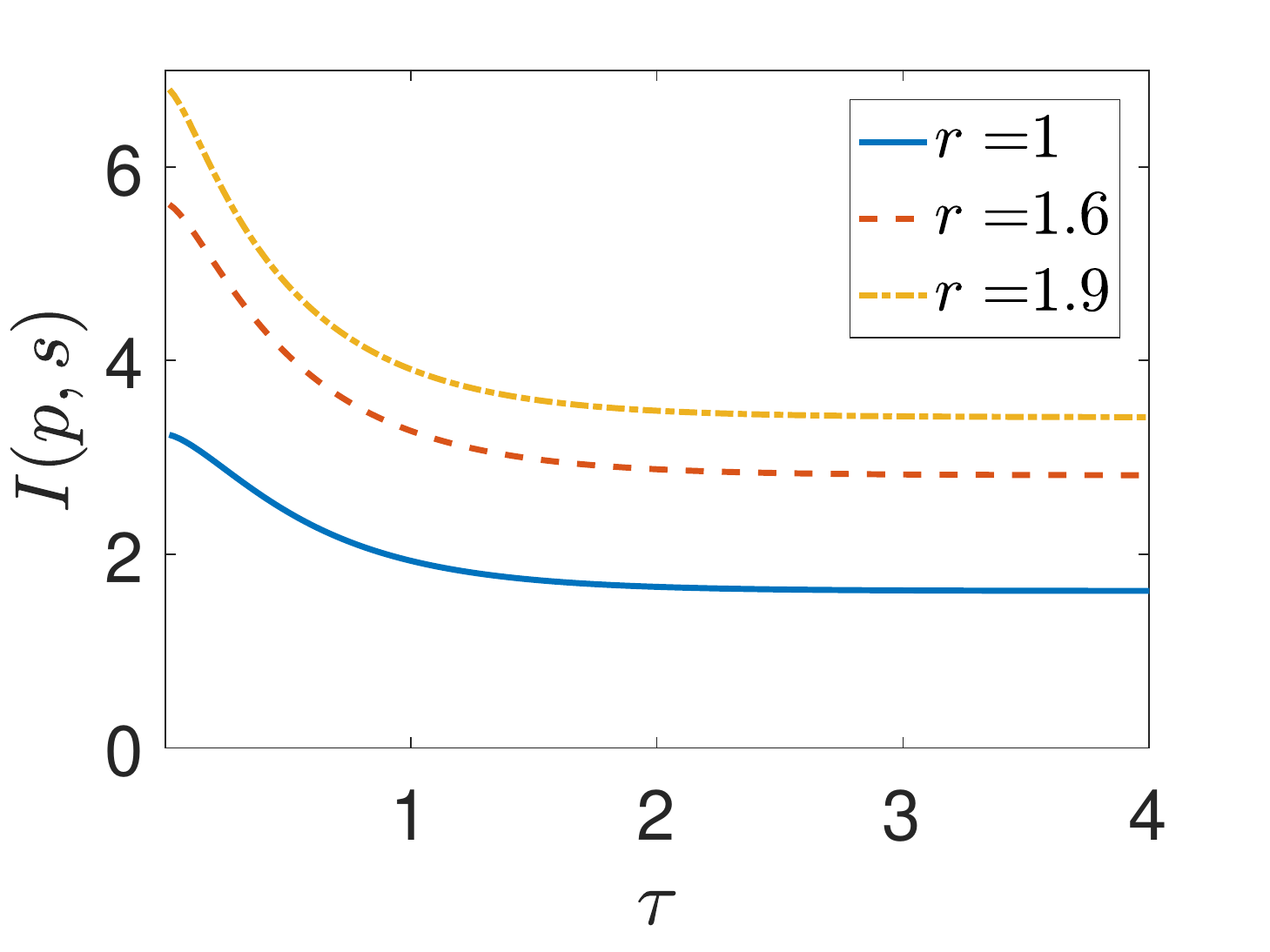}%
	}\hfill
	\caption{Three dimensional cavity and a non-equidistant spectrum with driving frequency $\Omega=\omega_{\bf s}+\omega_{\bf c}$. (a) The exponential creation of pairs of particles, as a function of time, in mode ${\bf s}$ for different values of the squeezing parameter $r$. (b) Entanglement degradation measured by the logarithmic negativity  as time evolves, vanishing for long times. (c) Mutual information as function of time. Even though quantum correlations disappear for long times, classical correlations among the modes persist in the long time limit.}
	\label{fig:suma}
\end{figure*}
\section{Entanglement degradation}\label{sec:Degrada}
In this section, we shall study the entanglement degradation in two cavities. Initially, both cavities are in an entangled two-mode squeezed state comprising mode ${\bf s}$ in cavity 1 and ${\bf p}$ in cavity 2. At $t=t_0$, we harmonically shake  cavity 1 (as seen in Sec. \ref{sec:DCE}),  keeping the distance between the mirrors fixed, and study how the entanglement between both cavities evolves in time (Fig. \ref{fig:esquema}). We can think of this system in the context of RQI as mimicking two observers in relative motion, one them static and the other moving back and forth harmonically, and that can only access only one mode (${\bf p}$ and ${\bf s}$ respectively).
 
The initial state of the system can be written in terms of the destruction operators as 
\begin{align}
	a_{\bf s}^{\text{in}}&=\cosh(r)a_{\bf s}+\sinh(r)a_{\bf p}^{\dagger}\nonumber\\
	a_{\bf p}^{\text{in}}&=\cosh(r)a_{\bf p}+\sinh(r)a_{\bf s}^{\dagger}
\end{align}
where $a_{\bf s}$ and $a_{\bf s}^\dagger$ are the destruction operators of the uncorrelated modes, satisfying $\langle a_{\bf j}^{\dagger} a_{\bf j} \rangle=0$ and $\langle a_{\bf j} a_{\bf j} \rangle=0$, for ${\bf j}={\bf s},{\bf p}$.
As has been mentioned before, there are two qualitatively different behaviors for the evolution of the field in the shaken cavity depending on whether the spectrum is approximately equidistant or unevenly spaced. We therefore analyze all cases separately.

\begin{figure*}
	\subfloat[\label{sfig:prod_vm}]{%
		\includegraphics[width=.33\linewidth]{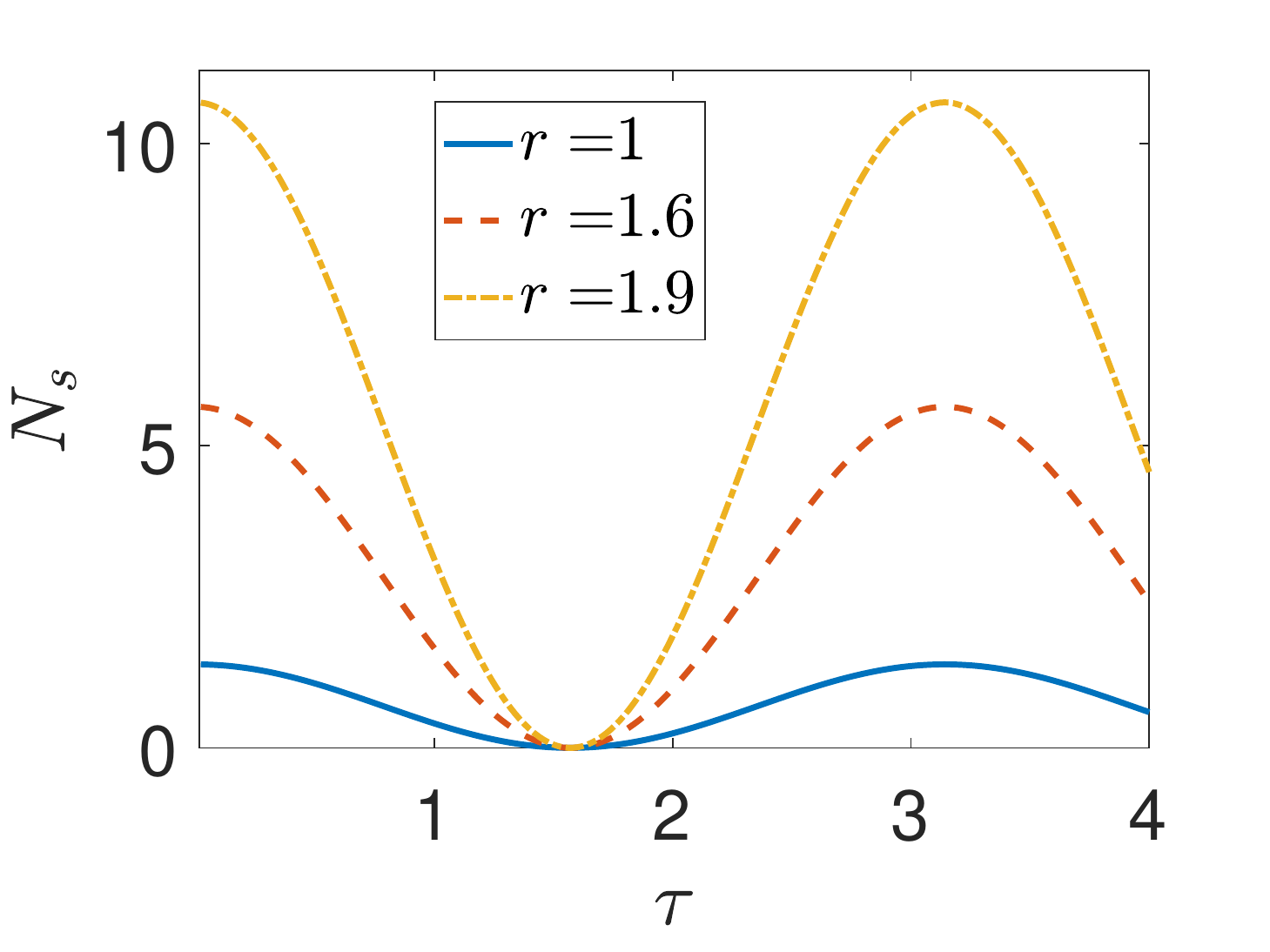}%
	}\hfill
	\subfloat[\label{sfig:prod_eta}]{%
		\includegraphics[width=.33\linewidth]{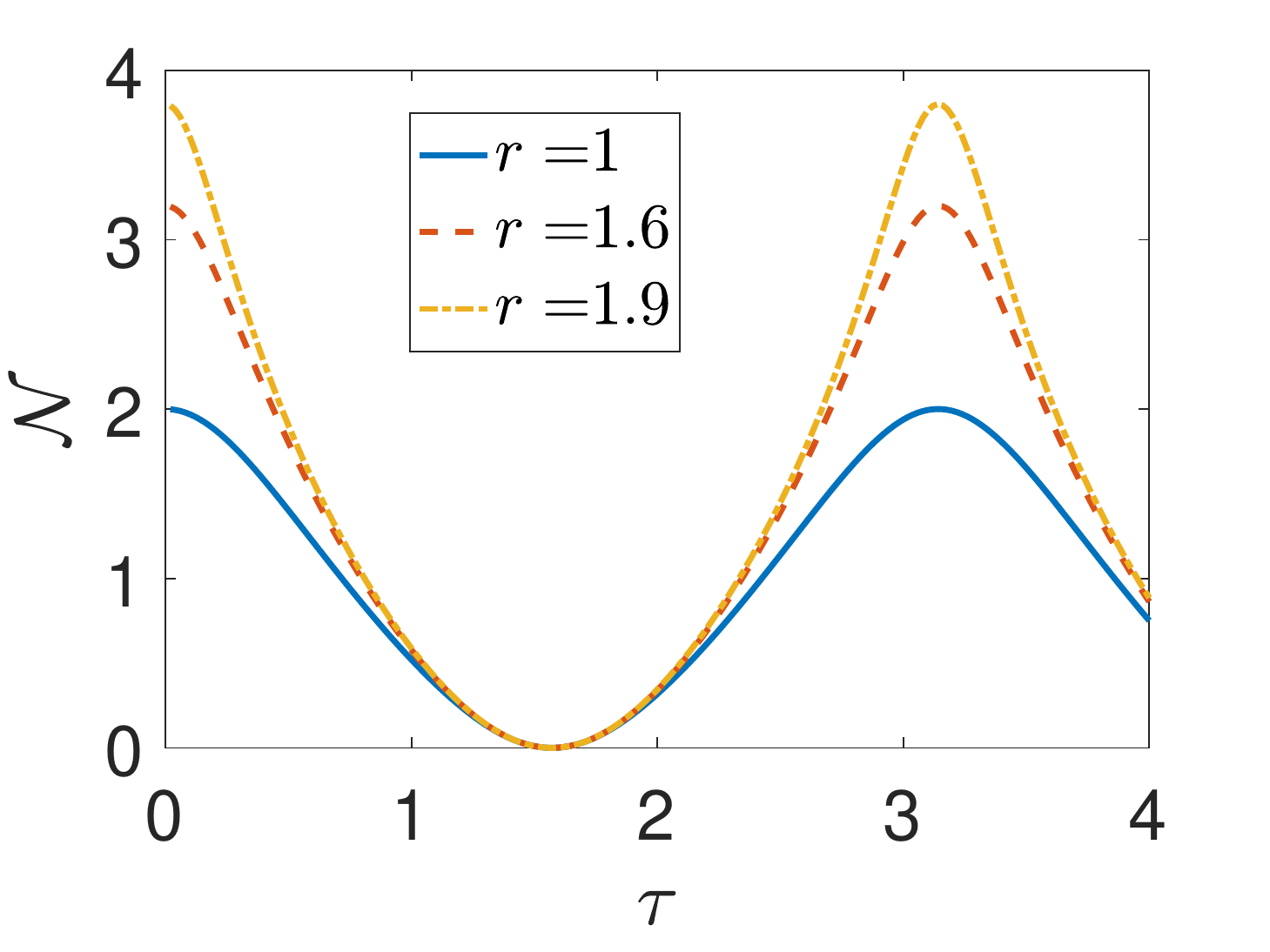}%
	}\hfill
	\subfloat[\label{sfig:prod_s_t}]{%
		\includegraphics[width=.33\linewidth]{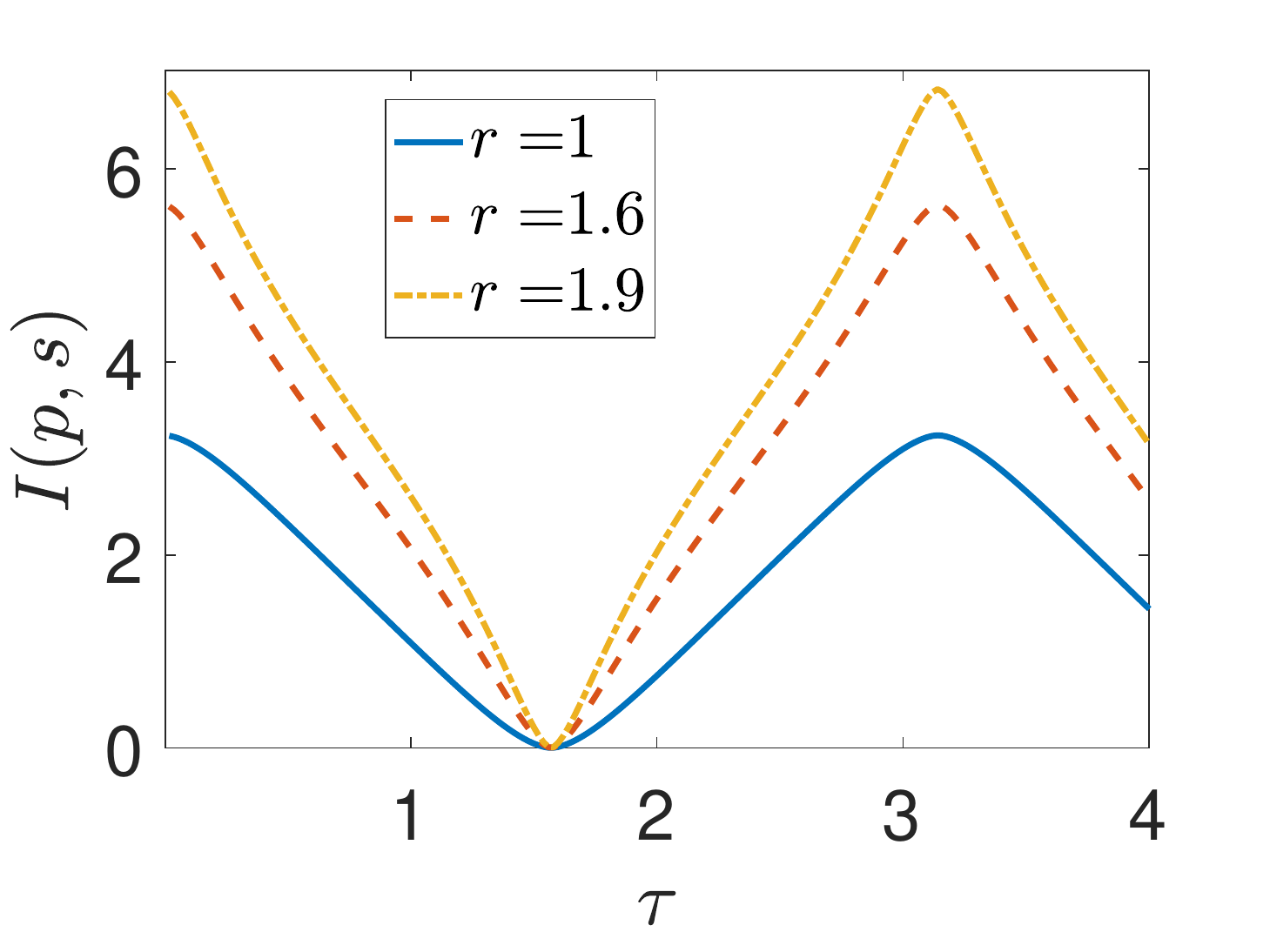}%
	}\hfill
	\caption{ Three dimensional cavity and a non-equidistant spectrum with driving frequency  $\Omega=|\omega_{\bf s}-\omega_{\bf c}|$.  In (a) The system behaves periodically with photons, originally in mode ${\bf s}$, switching back and forth between modes ${\bf s}$ and ${\bf c}$. This leads to an oscillatory behavior of the entanglement (b) and  the mutual information (c). They oscillate  between a maximum, when the photons are in ${\bf s}$ and zero, when they are in ${\bf c}$.}
	\label{fig:resta}
\end{figure*}


\subsection*{3D case and non-equidistant spectrum: Two coupled modes}


In the case that the shaking cavity is considered as three dimensional, the spectrum is not  evenly spaced. The excited mode ${\bf s}$ will then result coupled to at most one more mode, which we hereafter call mode ${\bf c}$.  As explained in the preceding Sec.\ref{sec:DCE},  the altered field state will result qualitatively different when the driving frequency is $\Omega=\omega_{\bf s}+\omega_{\bf c}$ or $\Omega=|\omega_{\bf s}-\omega_{\bf c}|$.\\


Firstly, we begin by analyzing the system  when  the cavity 1 is shaken with a frequency  $\Omega=\omega_{\bf s}+\omega_{\bf c}$. We further assume these two modes do not couple to any other mode. Shaking  cavity 1 with this frequency creates pairs of photons in modes ${\bf c}$ and ${\bf s}$. Accordingly to the derivations of the previous section, the new operators $a_{\bf s}^{\text{out}}$ and $a_{\bf c}^{\text{out}}$ are related to $a_{\bf s}^{\text{in}}$ and $a_{\bf c}^{\text{in}}$ by Eqs.(\ref{eq:sumas}) and (\ref{eq:sumap}). By considering the relation between creation operators and quadratures (\ref{eq:quadrature1}-\ref{eq:quadrature2}) we are able to calculate the covariance matrix, which in the basis $(q_{\bf p},p_{\bf p},q_{\bf s},p_{\bf s})$, is

\begin{widetext}
\begin{equation}
V_{{\bf p}|{\bf s}}=\frac{1}{2}\left|\begin{array}{cccc}
\cosh(2r) & 0 & \cosh(\gamma\tau)\sinh(2r) & 0\\
0 & \cosh(2r) & 0 & -\cosh(\gamma\tau)\sinh(2r)\\
\cosh(\gamma\tau)\sinh(2r) & 0 & (\cosh^2(r)\cosh(2\gamma\tau)+\sinh^2(r)) & 0\\
0 & -\cosh(\gamma\tau)\sinh(2r) & 0 &  (\cosh^2(r)\cosh(2\gamma\tau)+\sinh^2(r))
\end{array}\right|.
\end{equation} 
\end{widetext}
We can further  compute the long time properties of this state. We note that
\begin{align}
\Sigma=&\frac{1}{4}\cosh(2r)^2+\frac{1}{4}(\cosh^2(r)\cosh(2\gamma\tau)+\sinh^2(r))^2 \nonumber\\
&+\frac{1}{2}(\cosh(\gamma \tau)\sinh(2r))^2\xrightarrow[\tau \to\infty]{}\infty.
\end{align}
Before continuing, it is important to mention that the following limit remains finite
\begin{equation}
\frac{\det V_{{\bf p}|{\bf s}}}{\Sigma}\xrightarrow[\tau \to\infty]{}\frac{1}{4}.
\end{equation}
Hence, it is easy to see that  the negativity ($\mathcal{N}=-\log 2 \nu_- $) vanishes in the long time limit 
\begin{align}
\mathcal{N}=&-\log 2 \sqrt{\frac{\det V_{{\bf p}|{\bf s}}}{\Sigma}\frac{1}{\frac{1}{2}+\sqrt{\frac{1}{4}-\frac{1}{\Sigma}\frac{\det(V_{{\bf p}|{\bf s}})}{\Sigma}}}} \xrightarrow[\tau\to \infty]{} 0.
\end{align}

In addition, we can also study the mutual information between the cavities by calculating the symplectic eigenvalues of $2V_{{\bf p}|{\bf s}}$. They are 
 \begin{align}
 \eta_-&=1, \nonumber \\
 \eta_+&=|\cosh(r)^2\cosh(2\gamma \tau)-\sinh(r)^2|
 \end{align}
 and so the mutual information for long times is given by  
 \begin{align}
 I&=f(\cosh(r)^2\cosh(2\gamma \tau)+\sinh(r)^2) +f(\cosh(2r)) \nonumber\\
 &-f(|\cosh(r)^2\cosh(2\gamma \tau)-\sinh(r)^2|)\nonumber\\
 &\xrightarrow[\tau \to \infty]{} f(\cosh(2r)).
 \end{align}
In Fig.\ref{fig:suma} we compile all properties computed for this case. In panel (a), we show  the number of particles in mode ${\bf s}$. As can be seen, there is an exponential creation of pairs of photons in modes 
${\bf s}$ and ${\bf c}$ as  expected. In  panel (b), we can see the  time evolution of the entanglement as measured by the logarithmic negativity which is evidently degraded as time evolves,  vanishing for long times and destroying the initial entanglement between both cavities.  In panel (c),  we plot the mutual information that evidences that classical correlations among the modes persist for long times. This means that even though the entanglement vanishes, there are still classical correlations contained in the mutual information between the modes ${\bf p}$ and ${\bf s}$ in the long time limit.


We now consider the distinct case of  the cavity 1 being driven with a frequency  $\Omega=|\omega_{\bf s}-\omega_{\bf c}|$.
As mentioned before, this frequency does  not create new photons but only redistributes  the existing ones between the modes. In that case, the out operators are given by the Eqs.(\ref{eq:restas}) and (\ref{eq:restac}) and the covariance matrix can then be written as
\begin{widetext}
	\begin{equation}
	V_{{\bf p}|{\bf s}}=\frac{1}{2}\left|\begin{array}{cccc}
	\cosh(2r) & 0 & \cos(\gamma^+\tau)\sinh(2r) & 0\\
	0 & \cosh(2r) & 0 & -\cos(\gamma^+\tau)\sinh(2r)\\
	\cos(\gamma^+\tau)\sinh(2r) & 0 & (\cos^2(\gamma\tau)\cosh(2r)+\sin^2(\gamma^+\tau)) & 0\\
	0 & -\cos(\gamma^+\tau)\sinh(2r) & 0 &  (\cos^2(\gamma^+\tau)\cosh(2r)+\sin^2(\gamma^+\tau)) 
	\end{array}\right|. \nonumber
	\end{equation} 
\end{widetext}
The components of this matrix oscillate harmonically in time and for $\gamma^+\tau_n=\frac{\pi(2n+1)}{2}$ we obtain 
\begin{equation}
V_{{\bf p}|{\bf s}}=\frac{1}{2}\text{diag}(\cosh(2r),\cosh(2r),1,1)
\end{equation}
which means that the logarithmic negativity, the number of photons  in mode ${\bf s}$  and mutual information between ${\bf s}$ and ${\bf p}$ all vanish. Contrarily, for $\gamma^+\tau_m=\frac{m\pi}{\gamma}$, these magnitudes oscillate returning to their initial maximum values, as can be seen in Fig.\ref{fig:resta}.  In Fig.\ref{fig:resta}(a) we show the number of particles for mode ${\bf s}$. As can be inferred, there is no creation of new ones, but an oscillatory redistribution of the particles already in the cavity mode. Particles get transferred from mode ${\bf s}$ to mode ${\bf c}$, until no photons remain in ${\bf s}$ at  $\tau_m$.
In  Fig.\ref{fig:resta}(b) we show the entanglement temporal evolution which also results in a null entanglement at $\tau_m$. In Fig.\ref{fig:resta}(c), we show the temporal evolution of the mutual information between ${\bf s}$ in the first cavity and mode ${\bf p}$ in the second. \\


 
 \subsection*{1D case and equidistant spectrum: Infinite coupled modes}
 
 If the cavity  is one-dimensional then the spectrum is equidistant . This
means that  infinitely many modes get coupled  as the cavity is shaken (see Sec.\ref{sec:DCE}). In this situation the out operators are described by a more general  Bogoliubov transformation Eq.(\ref{eq:bogoliuvob}). Similarly as with the three dimensional cavities, we can see that there are two distinctly interesting cases to consider as for the driving frequency of cavity 1:  $\Omega=\omega_{1}$ and  $\Omega=q\omega_{1}$.\\

\begin{figure*}
	\subfloat[\label{sfig:prod_vm}]{%
		\includegraphics[width=.33\linewidth]{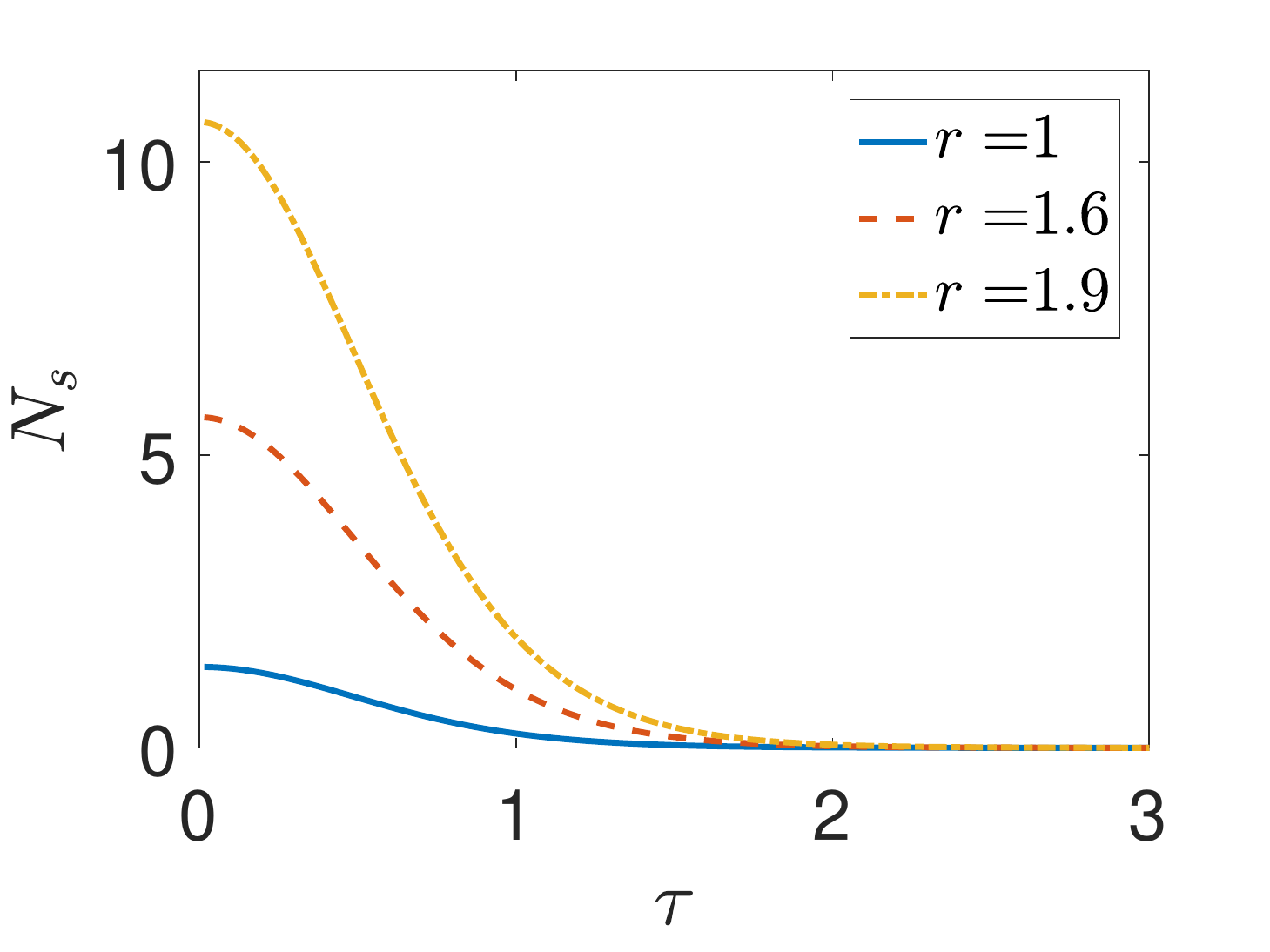}%
	}\hfill
	\subfloat[\label{sfig:prod_eta}]{%
		\includegraphics[width=.33\linewidth]{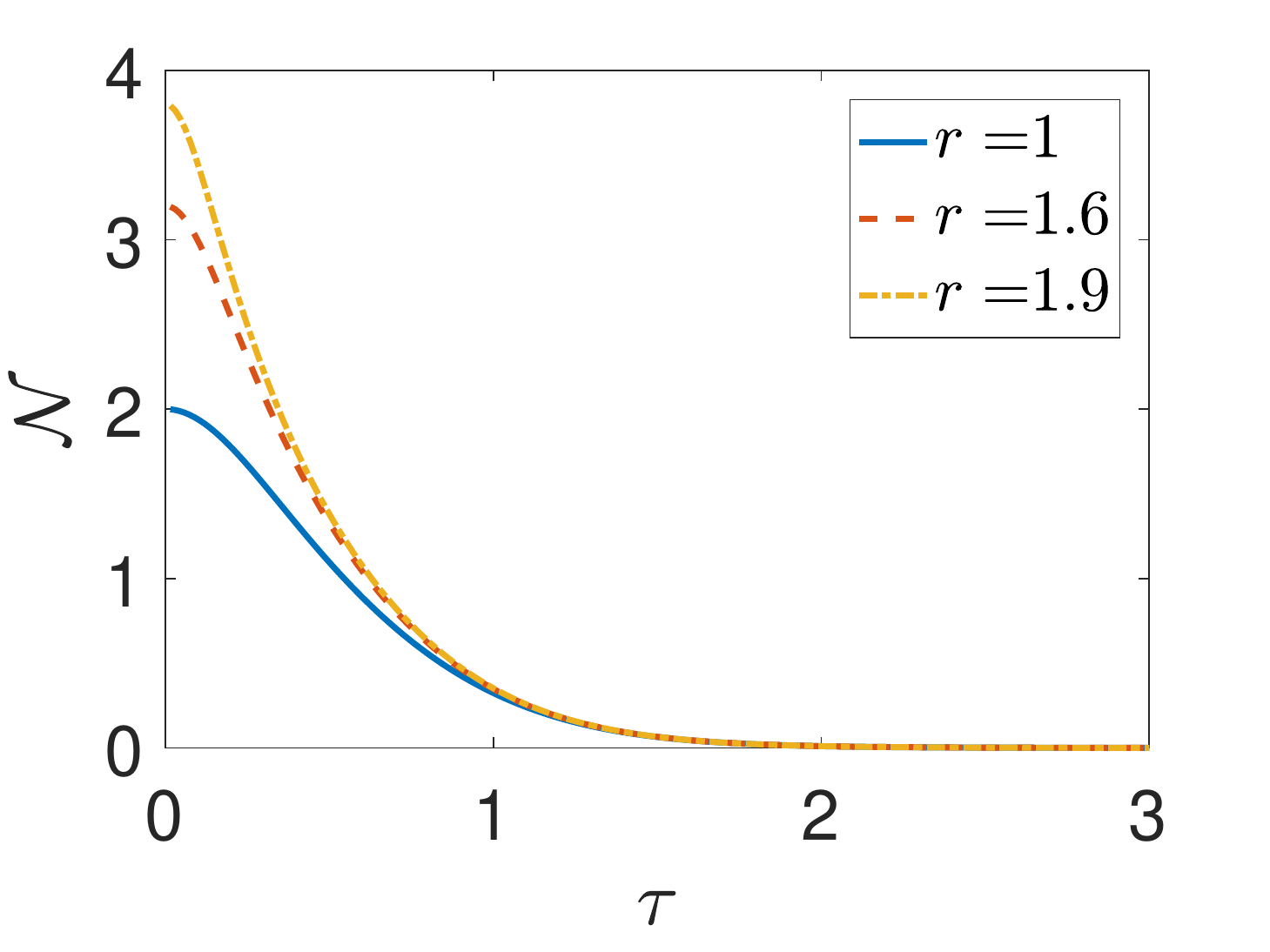}%
	}\hfill
	\subfloat[\label{sfig:prod_s_t}]{%
		\includegraphics[width=.33\linewidth]{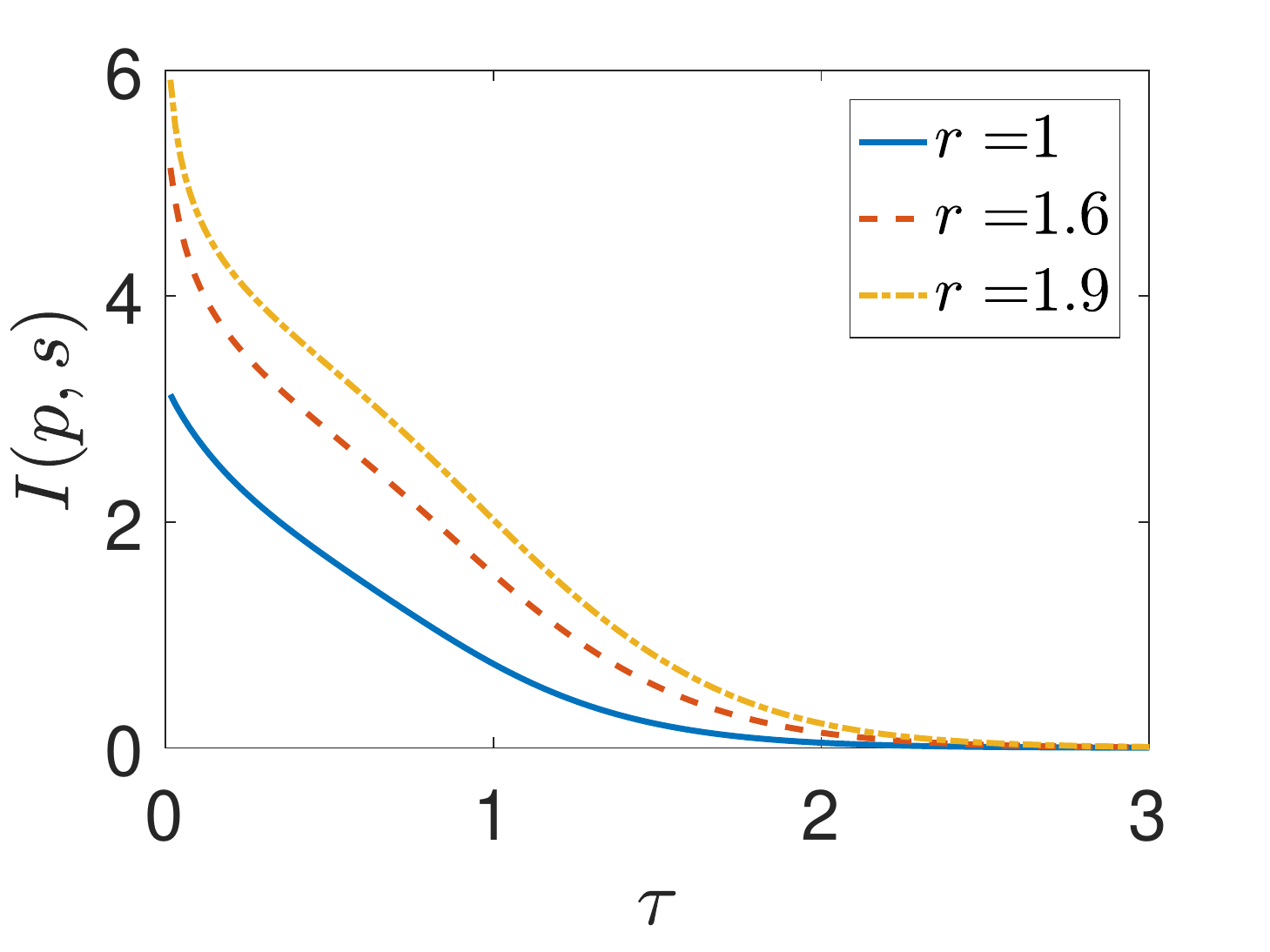}%
	}\hfill
	\caption{ One dimensional cavity and an equidistant spectrum with driving frequency  $\Omega=\omega_1$.  No creation of particles for this driving. (a) The particle behavior decreases as the initial photons in mode $s=1$ are lost to higher frequency modes for different values of the squeezing parameter $r$.  (b) This photon loss causes the entanglement and (c) mutual information  to vanish for long times.}
	\label{fig:1dw1}
\end{figure*}


Firstly, we start by considering $\Omega=\omega_1$. As it has already been pointed out, in this case there is no particle creation in  cavity 1.  Using Eq.(\ref{eq:bogoliuvob}), we can compute the covariance matrix for this case as
 
 \begin{widetext}
	\begin{equation*}
	V_{s|p}=\frac{1}{2}\left|\begin{array}{cccc}
	|\alpha_{ss}|^{2}\cosh(2r)+\sum_{j\neq s}|\alpha_{sj}|^{2} & 0 & \sinh(2r)\alpha_{ss} & 0\\
	0 & |\alpha_{ss}|^{2}\cosh(2r)+\sum_{j\neq s}|\alpha_{sj}|^{2} & 0 & -\sinh(2r)\alpha_{ss}\\
	\sinh(2r)\alpha_{ss} & 0 & \cosh(2r) & 0\\
	0 & -\sinh(2r)\alpha_{ss} & 0 &  \cosh(2r) 
	\end{array}\right|,
	\end{equation*} 
\end{widetext}
where we have already used that $\beta_{sj}=0$  and that $\alpha_{sj}(\tau)$,  is a real function given by Eq.(\ref{eq:coef1}). In order to understand the long time behavior of this state, we can use the Bogoliubov relation 
\begin{align}
1=\sum_{j=0}^\infty|\alpha_{sj}|^2-|\beta_{sj}|^2=\sum_{j=0}^\infty|\alpha_{sj}|^2
\end{align}
to reduce the problem in terms of  a unique Bogoliubov coefficient:
\begin{align}
(V_{s|p})_{11}&=(V_{s| p})_{22}=\frac{1}{2}\left(|\alpha_{ss}|^2\cosh(2r)+\sum_{j\neq s}^\infty|\alpha_{sj}|^2\right)\nonumber\\
&=\frac{1}{2}\left(|\alpha_{ss}|^2(\cosh(2r)-1)+1\right).
\end{align}
As a matter of fact, the explicit expression for this coefficient $|\alpha_{ss}|$ is already known to be \cite{Dodonov1996}
\begin{align}
	\alpha_{ss}(\tau)=&\sum_{j=1}^{s}\left[(s-1)!(s+j-1)!(-1)^{s-j}\right]\nonumber\\
	&\left[(s-1)!j!(s-j)!\right]^{-1}(\cosh\tau )^{-2j}\xrightarrow[\tau \to\infty]{}0. \nonumber 
\end{align}
All in all, the covariance matrix can be easily seen to converge to
\begin{equation}
V_{s|p}\xrightarrow[\tau \to\infty]{}\frac{1}{2}\text{diag}(1,1,\cosh(2r),\cosh(2r)).
\end{equation}
This means that not only the entanglement but  the mutual information as well vanish in the long time limit. 
In Fig.\ref{fig:1dw1}, we compile the results obtained when $\Omega=\omega_1$ in a one dimensional cavity behavior. In 
Fig.\ref{fig:1dw1}(a), we show the particle redistribution process for different initial values of the $r$ parameter. The number of particles in mode $s=1$ is lost to higher frequency modes since all cavity modes are coupled. The smaller the initial value of $r$, the sooner the particles are spread into other modes. In Fig.\ref{fig:1dw1}(b), we show that the entanglement evolution behaves accordingly to the number of particles $N_s$: the entanglement is lost very rapidly as time goes on and decrease asymptotically to zero.  In 
Fig.\ref{fig:1dw1}(c), the mutual information exhibits a similar behavior in concordance to the rest of the quantities considered.
It is important to note that the entanglement degradation in a one dimensional cavity is qualitatively different to that occurring in a three dimensional cavity.  In the former case, the loss of entanglement is due to the many cavity modes available after shaking the cavity. This implies a  redistribution of the existing particles in the cavity since no creation process takes place for this case. We can even think of a big environment to which information is lost. This is in contrast with the case of a three dimensional cavity where the entanglement is not lost asymptotically but rather oscillates following the particles.\\

 \begin{figure*}
 	\subfloat[\label{sfig:prod_vm}]{%
 		\includegraphics[width=.33\linewidth]{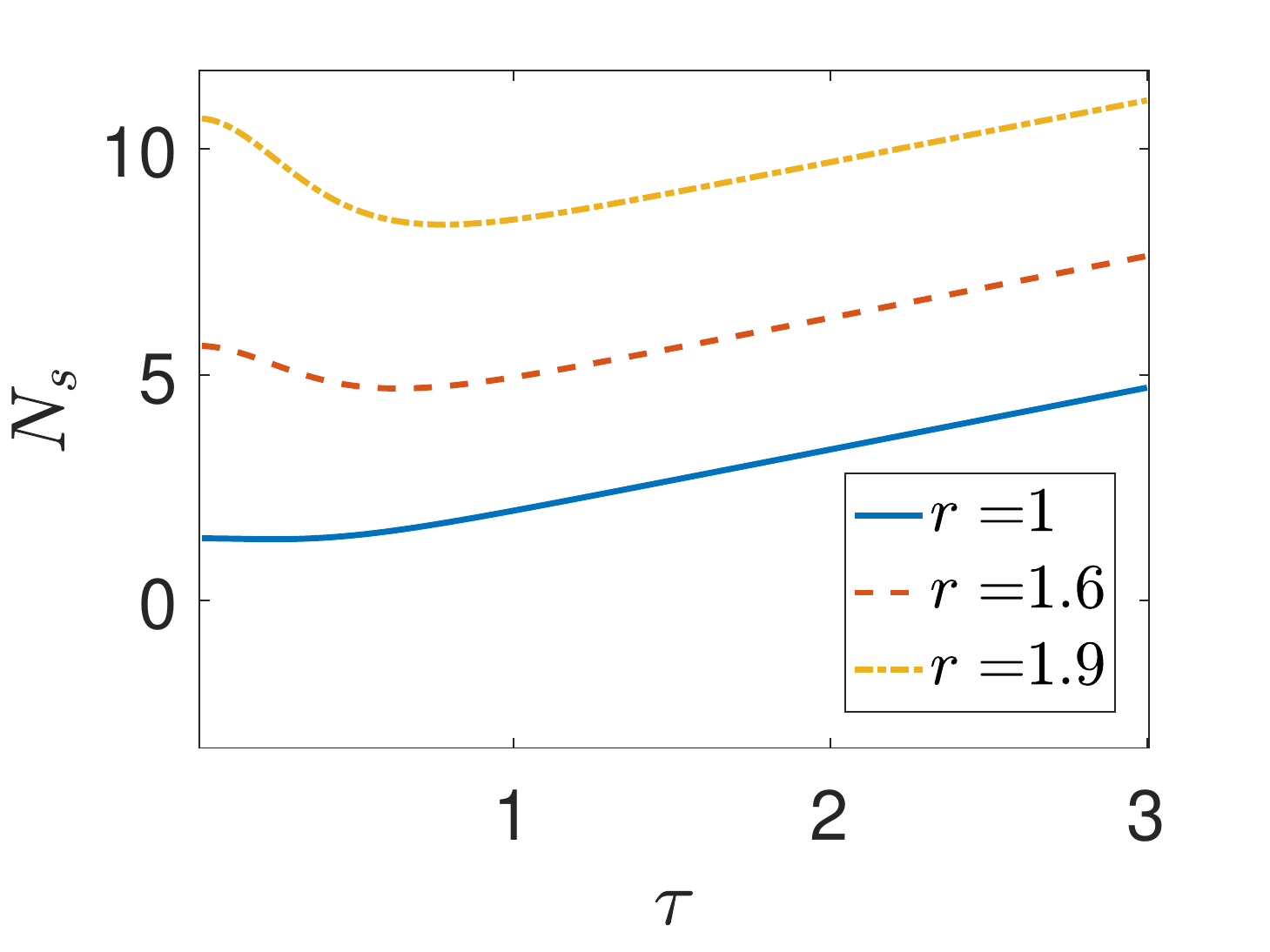}%
 	}\hfill
 	\subfloat[\label{sfig:prod_eta}]{%
 		\includegraphics[width=.33\linewidth]{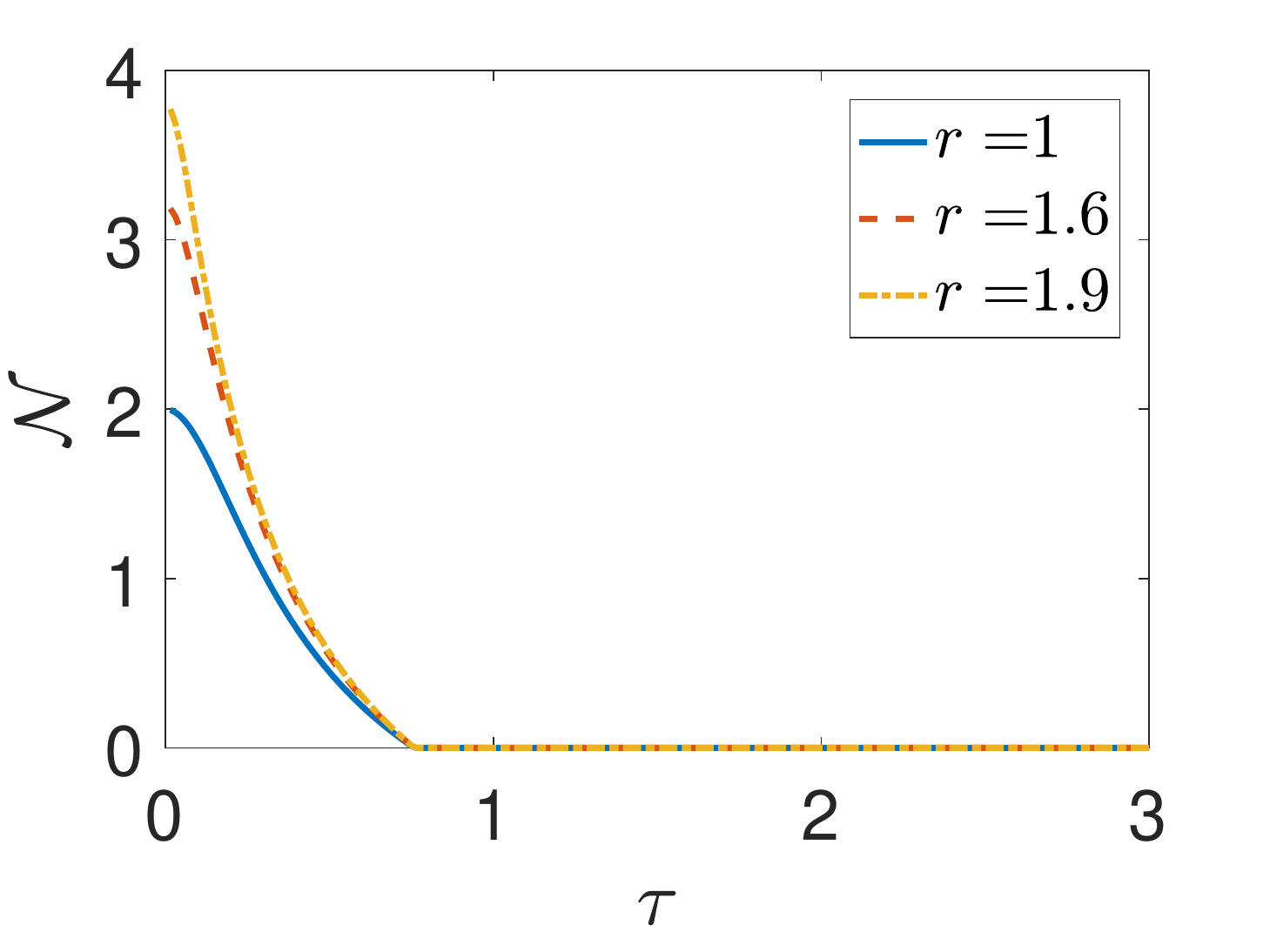}%
 	}\hfill
 	\subfloat[\label{sfig:prod_s_t}]{%
 		\includegraphics[width=.33\linewidth]{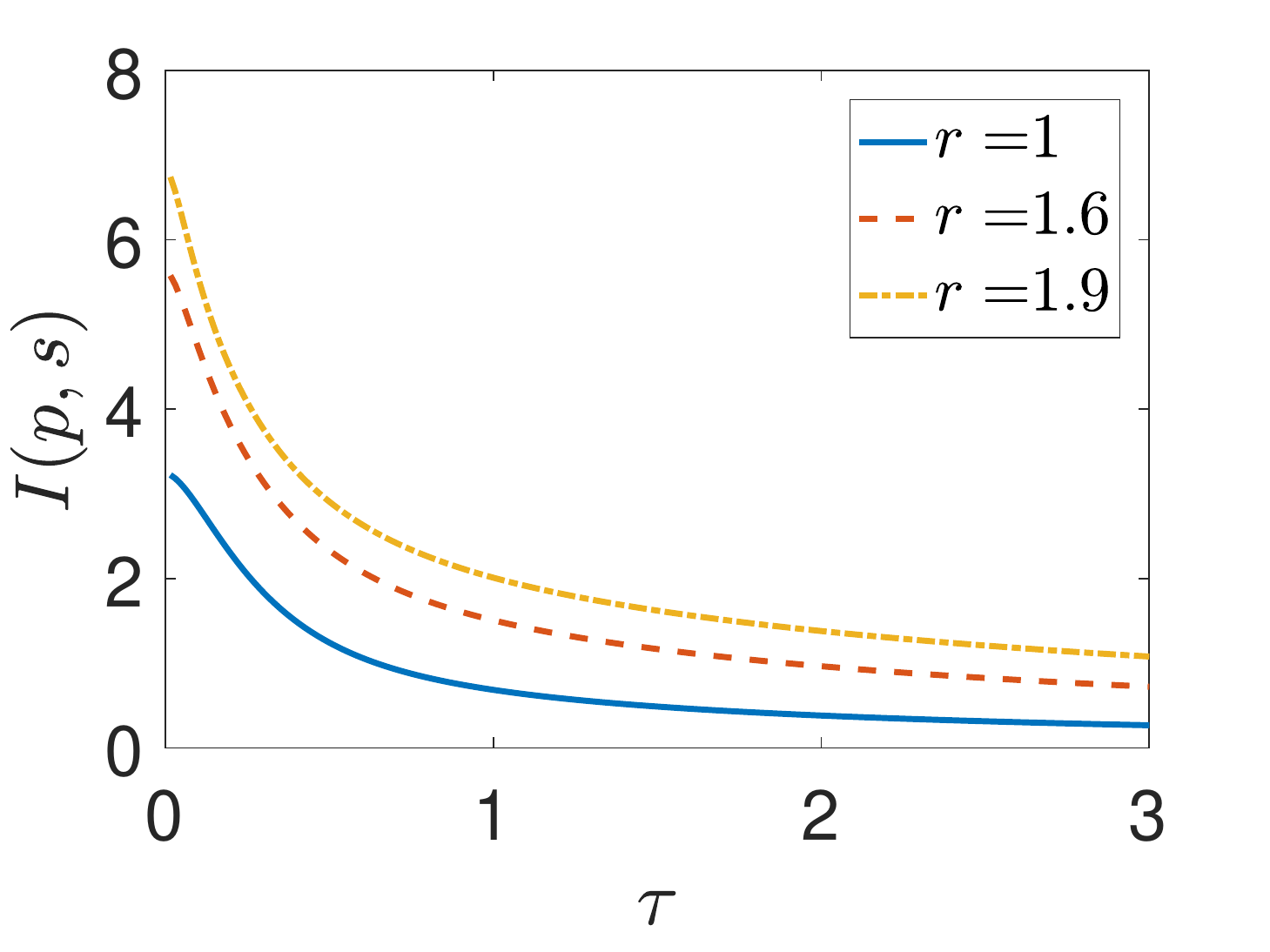}%
 	}\hfill
 	\caption{One dimensional cavity and an equidistant spectrum with driving frequency  $\Omega= q \omega_1$. (a) The number of photons initially decreases since a redistribution between the available modes takes place. As this case implies particle production, some time later the number of particles grow linearly as photon pairs are produced in modes $s$ and $c$. (b) The entanglement evidences a sudden death while (c) the mutual information behavior implies that classical correlations remain in the system  but approach zero for long times.  In this case we have assumed $q=3$.}
 	\label{fig:1dw3}
 \end{figure*}

Finally, we consider cavity 1 to be driven with a frequency $\Omega=q\omega_1$. This excites a parametric creation of photon pairs 
which is then redistributed along higher frequency modes. 
In this case the out operators are given by  Eq.(\ref{eq:1d}), from which we can calculate the covariance matrix as

 \begin{widetext}
	\begin{equation*}
	V_{s|p}=\frac{1}{2}\left|\begin{array}{cccc}
	|\alpha_{ss}|^{2}\cosh(2r)+\sum_{j\neq s}|\alpha_{sj}+\beta_{sj}|^{2} & 0 & \sinh(2r)\alpha_{ss} & 0\\
	0 & |\alpha_{ss}|^{2}\cosh(2r)+\sum_{j\neq s}|\alpha_{sj}-\beta_{sj}|^{2} & 0 & -\sinh(2r)\alpha_{ss}\\
	\sinh(2r)\alpha_{ss} & 0 & \cosh(2r) & 0\\
	0 & -\sinh(2r)\alpha_{ss} & 0 &  \cosh(2r) 
	\end{array}\right|.
	\end{equation*} 
\end{widetext}
We can simplify this expression by noticing that $V_{11}=V_{22}$. This is because $\beta_{1,k+np}=-\sqrt{1/(k+np)}\rho^{(k+np)}_{-1}=0$ for $k\neq p-1$ while $\alpha_{1,k+np}=\sqrt{1/(k+np)}\rho^{(k+np)}_{1}=0$ for $k\neq 1$ and so $\alpha_{1,k+np}\beta_{1,k+np}^*=0$, from which we conclude that $|\alpha_{1,k+np}\pm\beta_{1,k+np}|^2=|\alpha_{1,k+np}|^2+|\beta_{1,k+np}|^2$. By writing explicitly the covariance elements with Eq.(\ref{eq:covariance}), we can see that
\begin{align}
V_{11}+V_{22}=1+2N_{s},
\end{align}
where $N_{s}=\langle a_{s}^{\text{out}\dagger} a_{s}^{\text{out}} \rangle$ is the number of photons in mode $s$ and so $V_{11}=N_s+1/2$. The covariance matrix is then  reduced to only 3 independent components $V_{11},V_{33},V_{24}$. Therefore, we  proceed to study the asymptotic behavior of the state for long times. We must note that
\begin{equation}
\lim_{t\to \infty} V_{13}=- \lim_{t\to \infty}V_{24}<\infty , 
\label{eq:V13}
\end{equation}
since $V_{13},V_{24}\propto F(a,b,c;\kappa^2)$ and for $t\to \infty$ we have $\kappa \to 1$ and $F(a,b,c;\kappa^2)\to F(a,b,c;1)<\infty$. 

At this point, it is important to mention the fact that in ref \cite{Dodonov1998}, authors  have already proved that when a cavity with one oscillating mirror is driven with this frequency the number of photons from the vacuum $N_s(r=0)$ grows linearly with time \footnote{An exception for this behavior occurs when the frequency of the initially excited mode $s$ coincides with the shaking frequency $\omega_s=q\omega_{1}$. In this case the solutions behave as for $\Omega=\omega_1$, which we have already analyzed.}. Then using the equivalence between this setup and the shaken cavity, we see that
\begin{equation}
	V_{11}=\frac{1}{2}|\alpha_{ss}|^2(\cosh(2r)-1)+\frac{1}{2}+N_s(r=0) \\
	 \propto t.
	\label{eq:V11}
\end{equation}
 The previous simplifications allows us to write the explicit expressions as
\begin{align}
\det V&= V_{13}^2V_{24}^2-V_{11}V_{13}^2V_{33}-V_{11}V_{24}^2V_{33}+V_{11}^2V_{33}^2 \nonumber\\
\Sigma&= V_{11}^2-2V_{13}V_{24}+V_{33}^2.
\end{align}
Combining  with Eq.(\ref{eq:V13}) the long time limit yields
\begin{align}
\frac{\det V}{\Sigma}\xrightarrow[t\to \infty]{} V_{33}^2=(\frac{1}{2}\cosh(2r))^2.
\end{align}
This latter result implies that
\begin{align}
-\log 2 \nu_- &=-\log 2 \sqrt{\frac{\det V}{\Sigma}\frac{1}{\frac{1}{2}+\sqrt{\frac{1}{4}-\frac{1}{\Sigma}\frac{\det(V)}{\Sigma}}}}\\
&\xrightarrow[t\to \infty]{} -\log (\cosh(2r))<0,
\end{align}
leading to a surprising result:  the logarithmic negativity $\mathcal{N}=\max\{0,-\log 2 \nu_-\}$ manifests a sudden death at some finite time. As for the long time behavior of  the mutual information, we can use the symplectic eigenvalues  given by 
\begin{widetext}
	\begin{align}
	\eta_\pm&=\frac{1}{\sqrt{2}}\sqrt{V_{11}^2+V_{33}^2+2V_{13}V_{24}\pm\sqrt{V_{11}^4+V_{33}^4+4V_{33}^2V_{13}V_{24}-2V_{11}^2(V_{33}^2-2V_{13}V_{24})+4V_{11}V_{33}(V_{13}^2V_{24}^2)}} .
	\end{align}
\end{widetext}
By use of Eqs. (\ref{eq:V13}) and (\ref{eq:V11}), we have
\begin{align}
	\eta_+&\approx V_{11},\quad \text{for }\tau\gg1 \nonumber\\
	\eta_-&\approx V_{33},\quad \text{for }\tau\gg1
\end{align}
from which we conclude that the mutual information also vanishes in the long-time limit
\begin{align}
	I=f(V_{11})+f(V_{33})-f(\eta_-)-f(\eta_+) \xrightarrow[\tau \to \infty]{}0.
\end{align}

In Fig.\ref{fig:1dw3} we present the numerical results for $q=3$.  In Fig.\ref{fig:1dw3}(a), we compute the numerical evolution of $N_s$, while in Fig.\ref{fig:1dw3}(b), we show the entanglement degradation as time evolves.  It is easy to note  that the entanglement is qualitatively different from the previous cases as it dies suddenly in a finite time. This is due to the fact that two different effects are now combined and affect the entanglement: photon redistribution and pair creation. In Fig.\ref{fig:1dw3}(c), we can note that the mutual information, however, decreases slowly and asymptotically to zero.
This last case is completely different to all others. It implies many infinite modes coupled in a cavity in addition to particle creation process. Initially, it can be seen that the number of particles $N_s$ decreases because they are redistributed to higher modes available.  This is much evident for $r>1$. For a critical time the particle creation rate starts to gain importance and the number of particles in mode $s$ starts increasing. Surprisingly, this critical time is similar to the time the entanglement suddenly dies (or becomes zero). This combination of factors in not present in the other cases described above.

\begin{figure*}
	\subfloat[\label{sfig:2modos_logneg_sum}]{%
		\includegraphics[width=.4\linewidth]{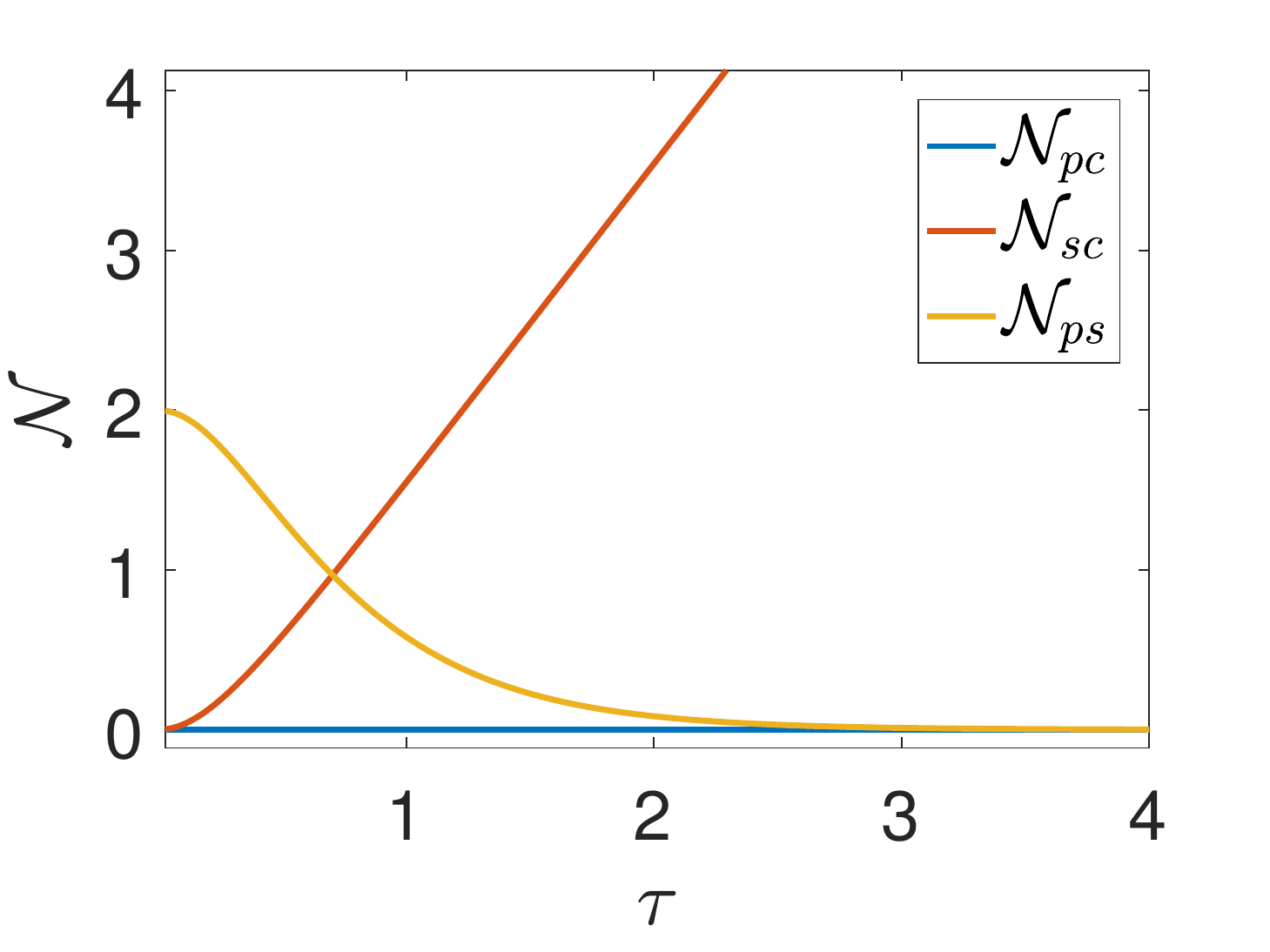}%
	}
	\subfloat[\label{sfig:2modos_logneg_dif}]{%
		\includegraphics[width=.4\linewidth]{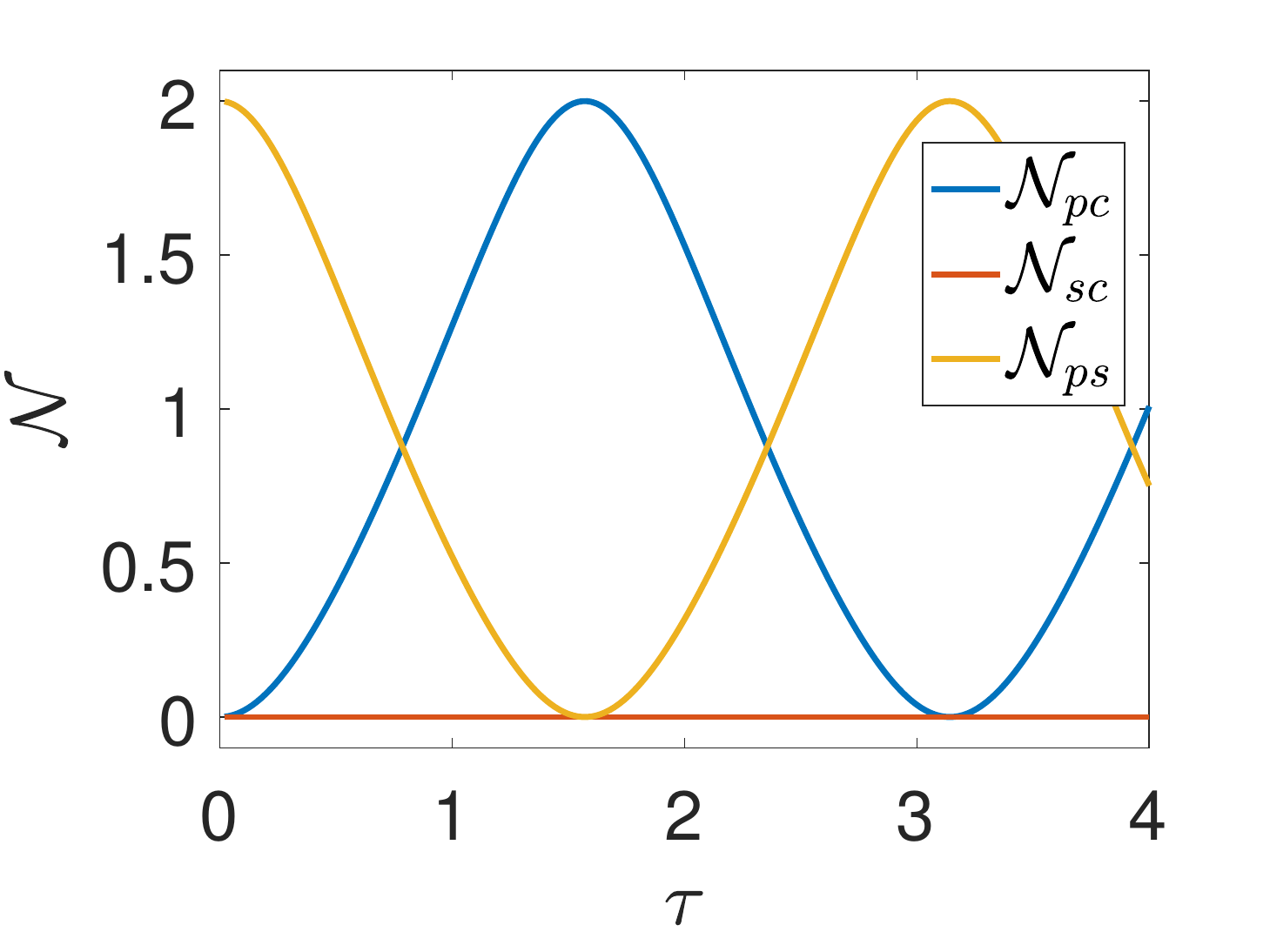}%
	}\hfill
	\caption{(a) Three dimensional cavity and a non-equidistant spectrum with driving frequency $\Omega=\omega_{\bf s}+\omega_{\bf c}$ and $r=1$. Entanglement $\cal N$ between modes $\bf p$ and $\bf c$, $\bf s$ and $\bf c$, $\bf p$ and $\bf s$ measured by the logarithmic negativity  as time evolves. The creation of photons pairs entangles modes $\bf s$ and $\bf c$ ($\mathcal{N}_{{\bf sc}}$), but it does not generate entanglement between $\bf p$ and $\bf c$ ($\mathcal{N}_{{\bf pc}}$); instead it degrades the entanglement $\mathcal{N}_{{\bf ps}}$. (b) Three dimensional cavity and a non-equidistant spectrum with driving frequency  $\Omega=|\omega_{\bf s}-\omega_{\bf c}|$. The system behaves periodically with photons, originally in mode ${\bf s}$, switching back and forth between modes ${\bf s}$ and ${\bf c}$. This oscillatory behavior also manifests in the entanglement dynamics $\cal N$ between modes of the first cavity with the second.}
	\label{fig:2modos_logneg}
\end{figure*}

\section{Entanglement Redistribution}\label{sec:Redis}
In this section, we shall study where the initial entanglement ${\cal N}_{\bf ps}$ (between cavity modes $\bf p$ and $\bf s$) goes after having altered the state of the field inside the cavity by means of the dynamical Casimir effect. It is important to note that in this manuscript we are studying a situation in which entanglement degradation exhibits different behaviors depending mainly on the spectrum of the cavity and the frequency of oscillation in consideration. As we have already seen that degradation takes place in significative different ways depending mainly on the cavity considered, in the following we shall study the entanglement dynamics among different cavity pairs available in the cavity.

For the three dimensional cavity, we have already mentioned that  only two modes will couple. The external excitation can then be considered as  $\Omega=|\omega_{\bf s} \pm \omega_{\bf c}|$,  leading to considerably different behaviors regarding creation of particles and entanglement degradation.
In the case that $\Omega=\omega_{\bf s} + \omega_{\bf c}$ we have shown  particle production is exponential in time. As the external frequency $\Omega$ excites modes $\bf s$ and $\bf c$,  particles of these modes are created in entangled pairs. The particles generated in mode $\bf c$ are never entangled with those of mode $\bf p$ but as time evolves they increase their entanglement with mode $\bf s$ while degrading the entanglement between $\bf p$ and $\bf s$ (see Fig. \ref{fig:2modos_logneg}(a)).

On the other hand, if $\Omega=|\omega_{\bf s} - \omega_{\bf c}|$, we have shown that there is no particle creation inside the cavity. This means that, as the cavity is shaken, the number of particles initially in mode $\bf s$ are transferred to mode $\bf c$ and then back to mode $\bf s$, exhibiting an oscillatory behavior. This is due to the fact that no extra photons are created, but only a redistribution of particles  in the cavity takes place. Entanglement dynamics behaves similarly with information being transferred between the two modes. Initial entanglement ${\cal N}_{{\bf ps}}$ decreases as ${\cal N}_{{\bf sc}}$ increases, showing an oscillatory behavior out of phase (Fig. \ref{fig:2modos_logneg}(b)). 
It is important to note that for both cases considered of the three-dimensional cavity, entanglement dynamics takes place between only two modes, due to the non-equidistant distribution of the cavity modes.
\begin{figure*}
	\subfloat[\label{sfig:w1_variosmodos_Nfot}]{%
		\includegraphics[width=.4\linewidth]{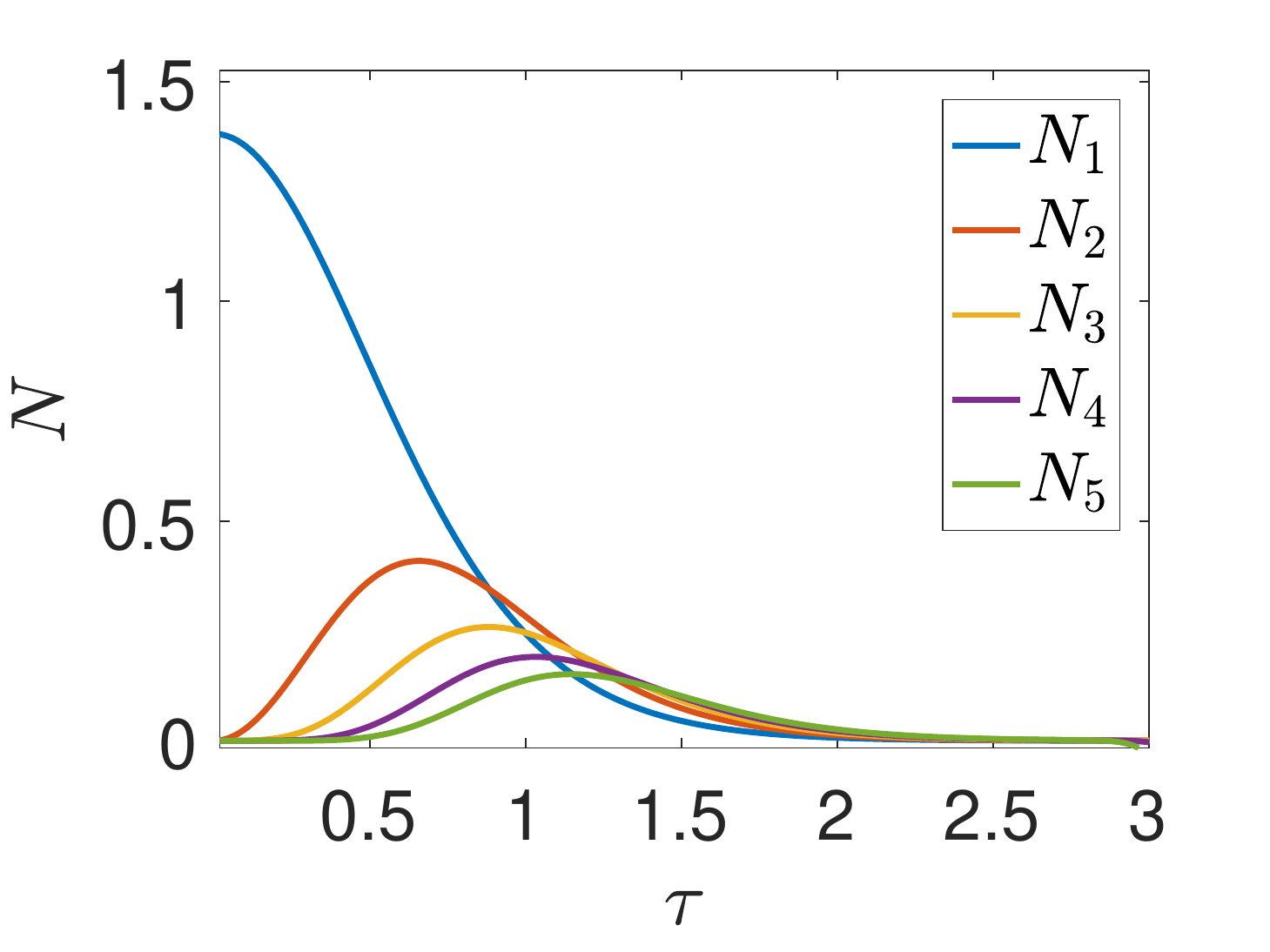}%
	}
	\subfloat[\label{sfig:w1_variosmodos_logneg}]{%
		\includegraphics[width=.4\linewidth]{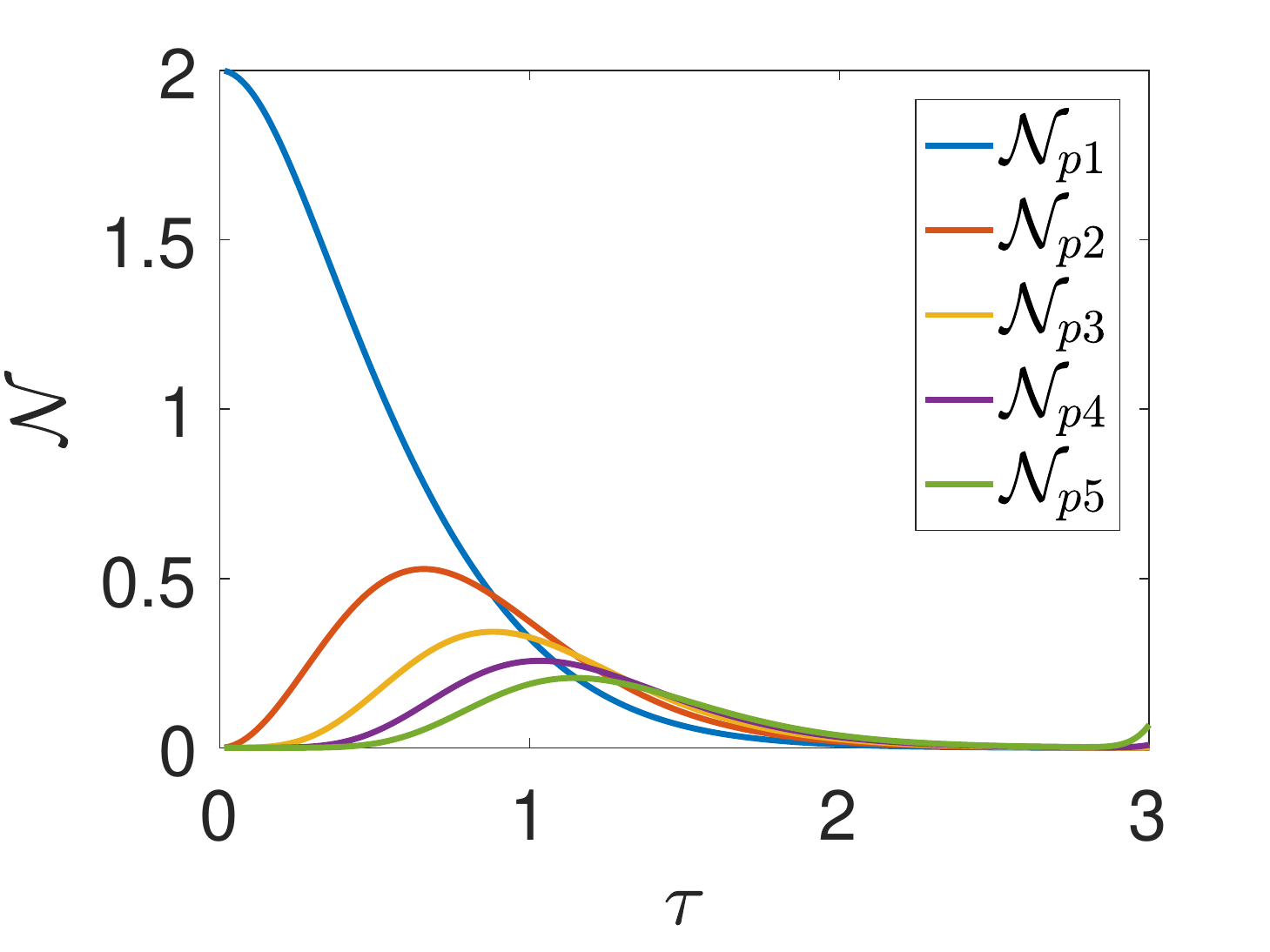}%
	}\hfill
	\caption{One dimensional cavity and an equidistant spectrum with driving frequency  $\Omega=\omega_1$, assuming $s=1$ and $r=1$. (a) Under these conditions no particles are produced but instead the initial particles in mode $1$ are redistributed towards higher frequency modes $2$, $3$,... This leads to an  entanglement flow to other higher cavities modes in detriment of  the initial entanglement between the modes of the moving and static cavities.}
	\label{fig:w1variosmodos}
\end{figure*}
\begin{figure*}
	\subfloat[\label{sfig:w3_variosmodos_Nfot}]{%
		\includegraphics[width=.4\linewidth]{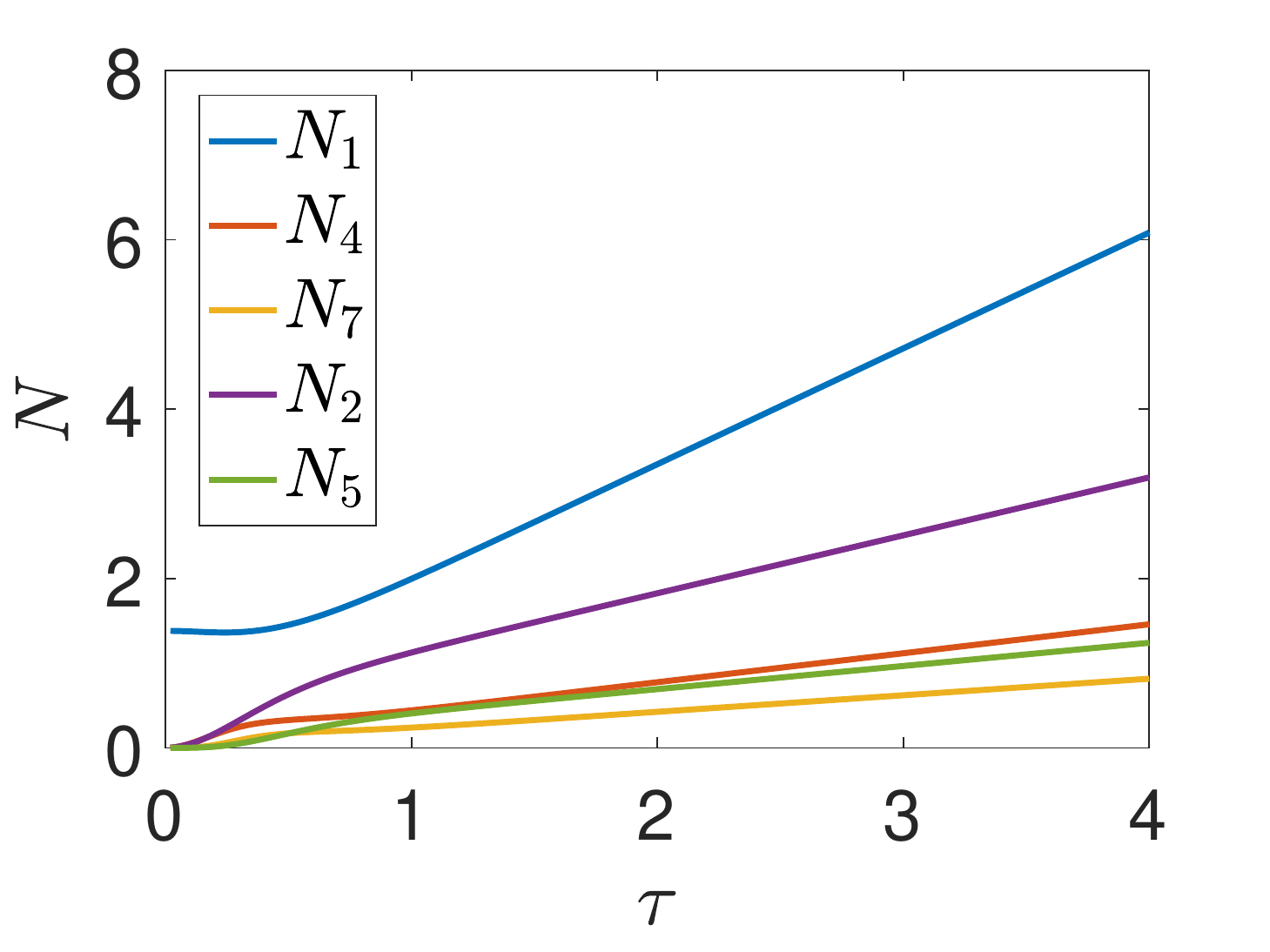}%
	}
	\subfloat[\label{sfig:w3_variosmodos_logneg}]{%
		\includegraphics[width=.4\linewidth]{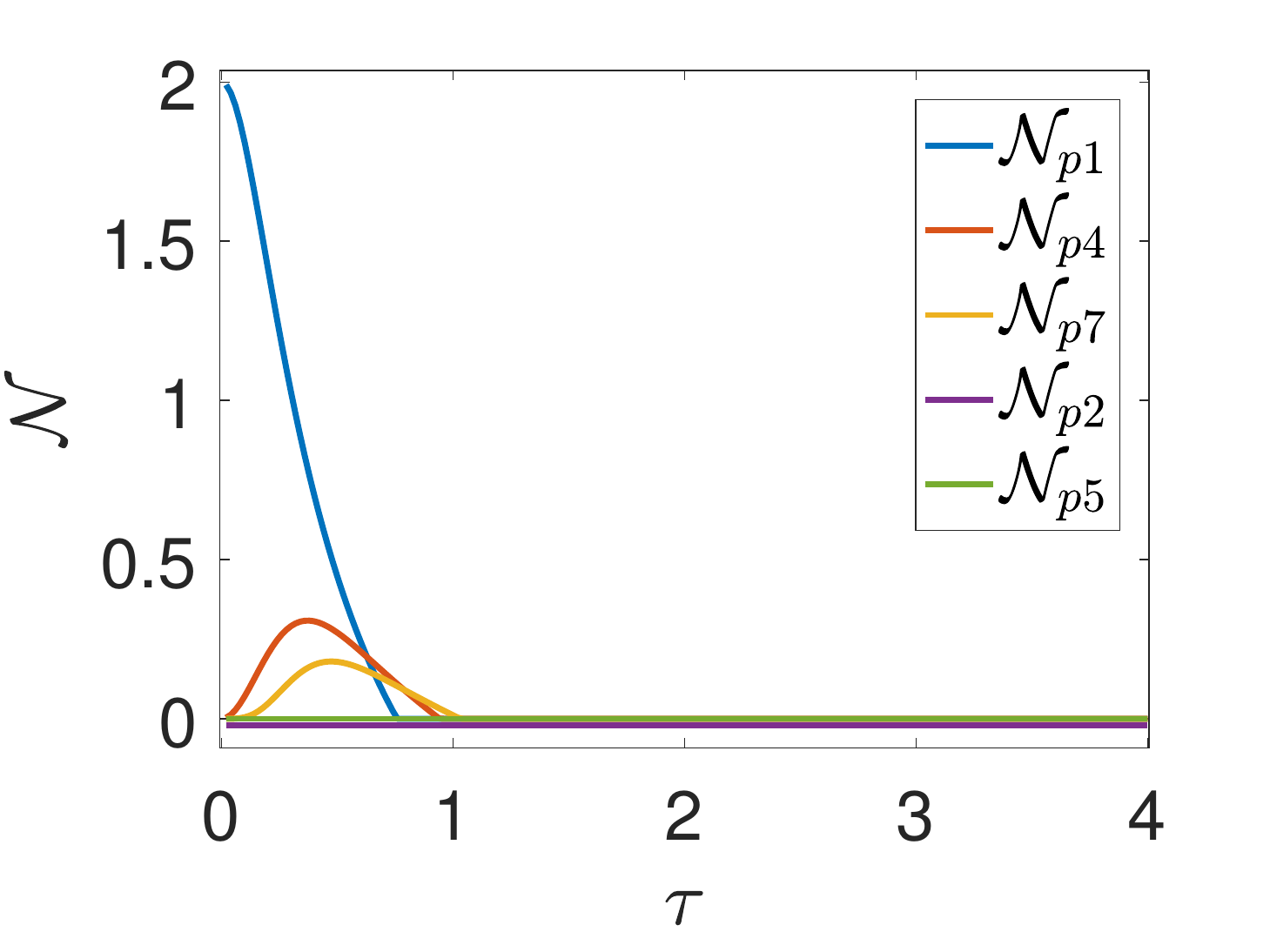}%
	}\hfill
	\caption{One dimensional cavity and an equidistant spectrum with driving frequency  $\Omega= q \omega_1$. In this case we have assumed $q=3$, $s=1$ and $r=1$. (a) The number of photons in mode $1$ initially decreases since a redistribution towards higher modes $4$, $7$, $10$, ... takes place. As this case also implies particle production, some time later the number of particles grows linearly as photon pairs are produced in modes $1$ and $2$. (b) The photons produced in modes $2$, $5$, $8$, ... are entangled with modes $1$, $4$, $7$, ... but not with mode $p$ in the static cavity, analogously to the three dimensional case Fig. \ref{fig:2modos_logneg}(a). However, entanglement that was originally between $1$ and $p$ is redistributed into  higher frequency modes of the form $1+3n$ with n a natural number and have a sudden death at finite time.}
	\label{fig:w3variosmodos}
\end{figure*}

In the case of a one-dimensional cavity, the resulting spectrum is equidistant, and cavity mode coupling takes place between many cavity modes. This implies a more complex dynamics, in which photon- and entanglement-transfer into higher frequency modes take place. When external driving is $\Omega=\omega_1$,
we have shown that there is no particle production inside the cavity. In Fig. \ref{fig:w1variosmodos}(a), we see the number of particles in different allowed cavity modes. As it can be noted, initially there are only particles in mode $s=1$. As time evolves, the number of particles  in mode $s=1$ decreases, as photons are transferred to higher cavity modes ($N_{s>1}$ temporarily increases). In Fig. \ref{fig:w1variosmodos}(b), we can note that as the number of particles in higher frequency modes increases, the initial entanglement is transferred to other pairs of cavity modes, leading to an increase of ${\cal N}_{p2}$, ${\cal N}_{p3}$ and so on. In this case, all cavity modes are allowed. The higher the frequency mode, the less degree of entanglement it receives. In the long time regime, the entanglement degradation is complete.

Finally, we must mention the case where $\Omega=q\omega_1$. Under this condition, there is particle production inside the cavity between pairs of modes whose frequency add up to $\Omega$ as can be seen in Fig.\ref{fig:w3variosmodos}(a). However, it is not the only consequence that we must take into the consideration. As the spectrum is equidistant, infinitely many modes can couple. Therefore, we  have shown that in this particular case, there is particle redistribution and creation inside the cavity. This leads to an entanglement dynamics different to all cases considered before. Initial entanglement  between modes $s$ and $p$ is redistributed towards higher frequency modes of the form: $\mod_q(s)+q\ n$ with $n$ a natural number and $\mod_q(s)$ the remainder of $s$ when divided by $q$ but eventually have a sudden death. However, the pairs produced between $\mod_q(s)$ and $q-\mod_q(s)$ do not entangle $q-\mod_q(s)$ with $p$ and therefore none of the higher frequency modes along which they are distributed (of the form $q-\mod_q(s)+q\ n$) get entangled with $p$ (see Fig. \ref{fig:w3variosmodos}(b) for an example).
The loss of entanglement is precisely explained as a redistribution of the inertial entanglement and the generation of multipartite quantum correlations among accessible and inaccessible modes inside the cavity.

\section{Conclusions}\label{sec:Conclu}

In this work we have studied how the entanglement and classical correlations between modes ${\bf s}$ and ${\bf p}$ in two different cavities are modified  when one of them is in relative oscillatory motion. In achieving so, we have firstly 
reviewed the dynamical Casimir Effect. We have presented the reigning equations for a two-moving wall (``shaker") cavity and computed the Bogoliubov transformation. Further, we have analyzed the particle creation process for a non-equidistant and an equidistant spectrum and stressed their similitudes and differences.  It is important to mention that even though the analysis of the DCE has been performed previously, the equivalence between a rigid translational movement of a two-wall moving cavity and a single moving one with twice the amplitude of movement is a new contribution which allowed to achieve a complete analytical description of the system.


We have  found that there are four qualitatively different behaviors for the system, depending on wether the spectrum is equidistant or not. The other feature that determines the behavior is whether the driving frequency $\Omega$ is able to create new photons or just redistribute the initial existing ones. If the spectrum is unevenly spaced and there is an additional mode ${\bf c}$ such that $\Omega=|\omega_{\bf s}-\omega_{\bf c}|$ then photons oscillate between ${\bf s}$ and ${\bf c}$. This causes the entanglement and the mutual information among the cavities to oscillate in time. However, if ${\bf c}$ has a frequency such that $\Omega=|\omega_{\bf s} + \omega_{\bf c}|$ then pairs of photons are created in these modes, this degrades the entanglement between the cavities which goes asymptotically to zero.  In spite of this, classical correlations persist as the mutual information converges to a positive value in the long-time limit. This result is analogous to the one found in \cite{Fuentes2005} and \cite{Adesso2007} where the entanglement between two observers is degraded as one them accelerates but classical correlations persist.
Our results show that the situation becomes qualitatively different when the spectrum of the cavity is evenly spaced, since this causes infinitely many modes the get coupled. In this case, if the moving cavity is shaken with its fundamental  frequency, no photons are produced. However, as time goes on the photons in the initially excited mode are eventually lost which causes the entanglement and mutual information between the cavities to vanish as well. On the other hand, if the cavity oscillates with a frequency that is an uneven harmonic, there is also  production of photons that forces  the entanglement to have a sudden death in a finite time, while the mutual information goes asymptotically to zero. This situation produces a similar result to what was found in \cite{Hu2012} where the entanglement between two harmonic oscillators was suddenly completely lost as one of them was accelerated.\\

Finally, by looking at what happens to other cavity modes, we have seen how this degradation occurs as a consequence of essentially two different processes which are the redistribution and pair creation of particles inside the moving cavity. In the case of an unevenly spaced spectrum and $\Omega=\omega_{\bf s}+\omega_{\bf c}$ the degradation is caused by particle creation and increases with time, while for  $\Omega=|\omega_{\bf s}-\omega_{\bf c}|$ it is caused by photon redistribution to a second mode and oscillates in time. On the other hand, for a cavity with an evenly spaced spectrum and shaken with its fundamental frequency the entanglement degradation stems solely from particle redistribution to higher frequency modes, while if it is shaken with an uneven harmonic of the fundamental frequency the degradation occurs as a combination of both effects. Finally, we have studied the entanglement redistribution so as to get an insight into where the entanglement is gone after altering the initial state of the system. We have shown that the answer relies on the particular particle creation process and coupling modes available in the cavity under each case considerated.\\

This setup captures many of the results previously found for observers and cavities in accelerated motion. In both cases we have a Bogoliubov transformations that generate photon pairs. However, in previous results of the existing literature, the alteration of the field state was due to  the Unruh Effect. In this work we exploit the non trivial structure of quantum vacuum and the effects derived from time dependent boundaries conditions. The result obtained
 is an apparent entanglement degradation and information loss for mostly cases considered.  We believe that this setup has  more promising experimental qualities since it relies on a bounded motion in an optomechanical system. While, there are still technological challenges that must be overcome for it to be tested in this exact setting, as the frequency of nanoresonators is not high enough, a simulation in a superconducting cavity is within experimental reach, since DCE has already been tested there.

\section*{Acknowledgements}
This work was supported by ANPCyT, CONICET, and Universidad de Buenos Aires; Argentina. 

\end{document}